\pdfoutput=1

\documentclass[format=acmsmall, review=false, screen=true, acmthm=true, authorversion=true, timestamp=false]{acmart}

\usepackage{booktabs} 

\usepackage{verbatim}
\usepackage{fancyhdr}
\usepackage[normalem]{ulem}
\usepackage{flushend}
\usepackage{float}
\usepackage{adjustbox}
\usepackage{epstopdf}
\usepackage{pgfplots}
\pgfplotsset{ytick style={draw=none},xtick style={draw=none}}
\usepackage{amssymb}
\usepackage{tikz, tkz-graph, ifthen }
\usepackage{listings}
\usepackage{minipage}
\usetikzlibrary{matrix,shapes,snakes,calc,decorations.pathreplacing,positioning}
\usepackage{tabularx,booktabs}
\usepackage{subfig}
\usepackage{graphicx}
\usepackage{pbox}
\usepackage{multirow,bigdelim}
\usepackage{makecell}
\usepackage{ctable}
\usepackage{color}
\usepackage{setspace}

\usepackage[title]{appendix}

\usepackage[utf8]{inputenc}

\newcommand\smtra{{\tt Z3RA}}
\newcommand\floor[1]{\lfloor#1\rfloor}
\newcommand\ceil[1]{\lceil#1\rceil}
\newcommand\low[1]{\underline{#1}}
\newcommand\high[1]{\overline{#1}}

\usepackage[ruled]{algorithm2e} 

\SetAlFnt{\small}
\SetAlCapFnt{\small}
\SetAlCapNameFnt{\small}
\SetAlCapHSkip{0pt}
\IncMargin{-\parindent}

\begin{document}

\title[Image Processing DSL and Precision Analysis]{Automatic Bitwidth Customization for 
Image Processing Pipelines on FPGAs through a DSL Compiler}

\author{Vinamra Benara}
\email{vinamra.benara@research.iiit.ac.in}
\author{Ziaul Choudhury}
\email{ziaul.c@research.iiit.ac.in}
\author{Suresh Purini}
\affiliation{%
 \institution{International Institute of Information Technology Hyderabad}
  \streetaddress{Gachibowli}
  \city{Hyderabad}
  \state{Telangana}
  \postcode{500032}
  \country{India}}
\email{suresh.purini@iiit.ac.in}
\author{Uday Bondhugula}
\affiliation{%
  \institution{Indian Insitute of Science Bangalore}
  \streetaddress{Department of Computer Science and Automation, Indian 
  Insitute of Science}
  \city{Bangalore}
  \state{Karnataka}
  \postcode{560012}
  \country{India}
}
\email{udayb@iisc.ac.in}


\begin{abstract}
High-level synthesis (HLS) has received significant attention in recent 
years for improving programmability of FPGAs. One could raise the level of
abstraction further by using domain-specific languages (DSLs), improving 
    productivity and performance simultaneously. PolyMage is a domain-specific
language and compiler for image processing pipelines. Its PolyMage-HLS backend
translates an input expressed as a DAG of image processing stages through the 
DSL into an equivalent circuit that can be synthesized on FPGAs, while 
    leveraging an HLS suite. 
    
The power and area savings while  performing arithmetic operations on 
fixed-point data type are well known to be  significant over using 
    floating-point data type. PolyMage-HLS stores  data at each stage of a 
    pipeline using a fixed-point data type $(\alpha, \beta)$ where $\alpha$ and 
    $\beta$ denote the number of  integral and fractional bits.  The integral 
    bitwidth ($\alpha$) requirement  at a pipeline stage can be inferred from 
    its range.  In this paper, we first propose an interval-arithmetic based 
    range  analysis algorithm to estimate the number of bits required to store
the integral part of the data at each stage of an  image processing pipeline.
The analysis  algorithm uses the homogeneity  of pixel signals at each stage to
cluster them and perform a combined  range analysis. Secondly, we propose a
software architecture for easily deploying
any kind of interval/affine arithmetic based range analyses in the DSL compiler.
Thirdly, we show that interval/affine arithmetic based techniques fail to take 
into account correlated computations across stages and hence could lead to poor 
    range estimates. These errors in range estimates accumulate across
stages, especially for iterative programs, such as Horn-Schunck Optical 
Flow, resulting in estimates nearly unusable in practice.  Then, we propose 
    a new range analysis technique using Satisfiability
Modulo Theory (SMT) solvers, and show that the range estimates obtained through 
it are very close to the lower bounds obtained through profile-driven analysis.
Finally, for estimating  fractional bitwidth ($\beta$) requirement at each stage 
of the pipeline, we propose a  simple
and practical heuristic search algorithm, which makes very few profile passes,
as opposed to techniques such as simulated annealing used in prior work.  The 
analysis algorithm attempts to minimize the number of fractional bits required
at each stage while preserving an  application-specific quality metric. We
evaluated our bitwidth analysis algorithms on four image processing  benchmarks
listed in the order of increasing complexity: Unsharp Mask, Down-Up Sampling, 
Harris Corner Detection and Horn-Schunck Optical Flow. The performance metrics
considered are quality, power and area. For example, on Optical Flow, the 
interval analysis based approach showed an 1.4$\times$ and 1.14$\times$ 
    improvement on area and power metrics over floating-point representation 
    respectively; whereas the SMT solver based approach showed 2.49$\times$ and 
    1.58$\times$ improvement on area and power metrics when compared to interval 
      analysis.
\end{abstract}

\maketitle

\definecolor{Code}{rgb}{0,0,0}
\definecolor{Decorators}{rgb}{0.5,0.5,0.5}
\definecolor{Numbers}{rgb}{0.5,0,0}
\definecolor{MatchingBrackets}{rgb}{0.25,0.5,0.5}
\definecolor{Keywords}{rgb}{0,0,1}
\definecolor{self}{rgb}{0,0,0}
\definecolor{Strings}{rgb}{0,0.63,0}
\definecolor{Comments}{rgb}{0,0.63,1}
\definecolor{Backquotes}{rgb}{0,0,0}
\definecolor{Classname}{rgb}{0,0,0}
\definecolor{FunctionName}{rgb}{0,0,0}
\definecolor{Operators}{rgb}{0,0,0}
\definecolor{Background}{rgb}{0.98,0.98,0.98}

\usetikzlibrary{shapes,arrows}
\usetikzlibrary{matrix, positioning, fit}
\definecolor{forestgreen}{rgb}{0.0, 0.5, 0.0}

\tikzstyle{block} = [draw=black, rounded corners, thick, line width=0.3mm, fill=blue!20, rectangle, minimum height=2em, minimum width=2em]
\tikzstyle{group} = [draw=black!100, dashed, line width=0.2mm, rectangle,
rounded corners, minimum height=1em, minimum width=1em, inner sep=0.4mm]
\tikzstyle{textblock} = [draw, fill=blue!20, rectangle, rounded corners]
\tikzstyle{input} = [draw, circle, minimum size=1pt]
\tikzstyle{output} = [draw, fill=blue!20, circle, minimum size=1pt]
\tikzstyle{pinstyle} = [pin edge={to-,thin,black}]

\tikzstyle{fancynodelong} =[draw=black!60,text=black, rounded corners, minimum height=2em, minimum width=4.3em]
\tikzstyle{fancynodesqr} =[fill=red!100, text=white, rounded corners, minimum height=4em, minimum width=4em]
\tikzstyle{fancycircle} =[draw,text=black, circle, minimum size=2em]

\tikzstyle{flowarrow} = [black, opacity=0.75]
\tikzstyle{blueflow} = [black, thick, opacity=0.8]
\tikzstyle{thickflowarrow} = [rounded corners, color=black, line width=0.5mm]

\tikzstyle{whitebox} =[fill=white!100, text=black, minimum height = 0.1mm, minimum width = 0.1mm]

\newcommand{\newlinebuf}[0]{
    \begin{tikzpicture}[auto, node distance=1.25cm,>=latex', scale=1, transform shape]

        \node [name = stencil_in] {
            \includegraphics[width=2.1cm]{images/stencil_in.png}
        };

        \node [below = 3cm of stencil_in, name = stencil_out] {
            \includegraphics[width=0.7cm]{images/stencil_out.png}
        };

        \node [fancycircle, fill=blue!100, below = 1.2cm of stencil_in, name=func] {\textbf{f}};

        \node [above left = -3.1cm and 0.3cm of stencil_in, name=inpic] {
            \includegraphics[width=6.5cm]{images/linebuf_in.png}
        };

        \node [below = 0.1cm of inpic, name=outpic] {
            \includegraphics[width=6.5cm]{images/linebuf_out.png}
        };

        \node [below = 0cm of stencil_in, name=brace] {
            \includegraphics[width=2.1cm]{images/brace.png}
        };

        \draw [<-, flowarrow] (func) to [out=90, in=-90] ++(0, 0.9);
        \draw [->, flowarrow] (func) to [out=-90, in=90] ++(0, -1.6);

        \draw [<-, thickflowarrow, color=blue!70, line width = 0.6mm] (stencil_in) to [out=90, in=90] ++(-5.1, 0.2);
        \draw [->, thickflowarrow, line width = 0.6mm] (stencil_out) to [out=-110, in=0] ++(-4.9, 0.3);

    \end{tikzpicture}
}

\newcommand{\naivebuffer}[0]{
    \begin{tikzpicture}
        \tikzset{VertexStyle/.append style = {
                minimum size = 2cm,font = \fontsize{17}{17}\bfseries},
                EdgeStyle/.append style = {->,>=stealth'},
                LabelStyle/.style = {above=2pt, text = blue!20!black, font = \fontsize{12}{12}\bfseries }}
        \SetGraphUnit{3.5}
        \Vertices{line}{img,blurx,blury,sharpen,mask}

        \Edge[label = 4w+1](img)(blurx)
        \Edge[label = 5](blurx)(blury)
        \Edge[label = 0](blury)(sharpen)
        \Edge[label = img:0,style={bend left, dashed}](sharpen)(mask)
        \Edge[label = 0](sharpen)(mask)
        \Edge[label = img:0,style= {bend left, dashed}](blurx)(sharpen)
        \tikzset{LabelStyle/.append style = {below=2pt}}
        \Edge[label = blury:0,style= {bend right, dashed}](sharpen)(mask)
    \end{tikzpicture}
}

\newcommand{\finalbuffer}[0]{
    \begin{tikzpicture}
        \tikzset{VertexStyle/.append style = {
                minimum size = 2cm,font = \fontsize{17}{17}\bfseries},
                EdgeStyle/.append style = {->,>=stealth'},
                LabelStyle/.style = {above=2pt, text = blue!20!black, font = \fontsize{12}{12}\bfseries }}
        \SetGraphUnit{3.5}
        \Vertices{line}{img,blurx,blury,sharpen,mask}

        \Edge[label = 4w+1](img)(blurx)
        \Edge[label = 5](blurx)(blury)
        \Edge[label = 0](blury)(sharpen)
        \Edge[label = img:0,style={bend left, dashed}](sharpen)(mask)
        \Edge[label = 0](sharpen)(mask)
        \Edge[label = img:5,style= {bend left, dashed}](blurx)(sharpen)
        \tikzset{LabelStyle/.append style = {below=2pt}}
        \Edge[label = blury:0,style= {bend right, dashed}](sharpen)(mask)
    \end{tikzpicture}
}

\newcommand{\dagdiamond}[0]{
    \begin{tikzpicture}
        \tikzset{VertexStyle/.append style = {
            minimum size = 1cm,font = \fontsize{12}{12}\bfseries},
            EdgeStyle/.append style = {->,>=stealth'}}
        \SetGraphUnit{2}
        \Vertex{s1} 
        \NOEA[unit=1.5](s1){s2}
        \SOEA[unit=1.5](s1){s3}
        \SOEA[unit=1.5](s2){s4}
        \Edge(s1)(s2)
        \Edge(s1)(s3)
        \Edge(s2)(s4)
        \Edge(s3)(s4)
    \end{tikzpicture}
}

\newcommand{\dagv}[0]{
    \begin{tikzpicture}
        \tikzset{VertexStyle/.append style = {
            minimum size = 1cm,font = \fontsize{12}{12}\bfseries},
            EdgeStyle/.append style = {->,>=stealth'}}
        \SetGraphUnit{2}
        \Vertex{s1} 
        \EA[unit=2](s1){s2}
        \SOEA[unit=1.5](s2){s5}
        \SOWE[unit=1.5](s5){s4}
        \WE[unit=2](s4){s3}
        \Edge(s1)(s2)
        \Edge(s3)(s4)
        \Edge(s2)(s5)
        \Edge(s4)(s5)
    \end{tikzpicture}
}

\newcommand{\dagvlabeled}[0]{
    \begin{tikzpicture}
        \tikzset{VertexStyle/.append style = {
            minimum size = 1cm,font = \fontsize{12}{12}\bfseries},
            EdgeStyle/.append style = {->,>=stealth'}}
        \SetGraphUnit{2}
        \Vertex{s1} 
        \EA[unit=2](s1){s2}
        \SOEA[unit=1.5](s2){s5}
        \SOWE[unit=1.5](s5){s4}
        \WE[unit=2](s4){s3}
        \Edge[label = 1,labelstyle={above=2pt}](s1)(s2)
        \Edge[label = 1,labelstyle={below=2pt}](s3)(s4)
        \Edge[label = 1,labelstyle={sloped, above=2pt}](s2)(s5)
        \Edge[label = w,labelstyle={sloped,below=2pt}](s4)(s5)
    \end{tikzpicture}
}

\newcommand{\overview}[0]{
    \begin{tikzpicture}[auto, node distance = 1cm, scale = 0.8, transform shape, font = \bf \scriptsize \sffamily]

        \node[whitebox, text width = 4.5em, name = a1, font = \normalsize ]
        {\textbf{PolyMage DSL code}};

        \node[fancynodelong, minimum height = 4em, minimum width = 5em, text 
        width = 5em, align = center, font = \normalsize, right = 0.7cm of 
        a1, name = a2]{\textbf{PolyMage-HLS compiler}};

        \node[fancynodelong, minimum height = 4em, minimum width = 5em, text width = 4em, align = center, font = \normalsize, right = 0.8 cm of a2, name = a3]{\textbf{Vivado HLS}};
        \node[fancynodelong, minimum height = 4em, minimum width = 4em, text width = 4em, align = center, font = \normalsize, right = 0.8cm of a3, name = a4]{\textbf{Vivado Design Suite}};
        \node[right = 1.4cm of a4, name = a5]{\includegraphics[width=1.4cm]{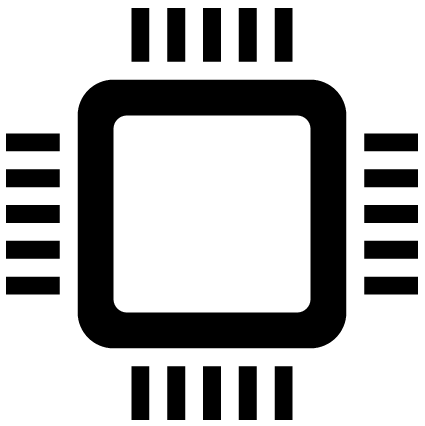}};

        \draw [->, line width = 0.3mm,black!60] (a1) -- (a2);

        \draw [->, line width = 0.3mm,black!60] (a2) -- (a3);
        \node [whitebox, above right = -0.6cm and 0cm of a2, minimum size = 0.4em, text width = 0.1em, name = a6_a7]{\textbf{C++}};

        \draw [->, line width = 0.3mm,black!60] (a3) -- (a4);
        \node [whitebox, above right = -0.6cm and 0cm of a3, minimum size = 0.4em, name = a7_a8]{\textbf{HDL}};

        \draw [->, line width = 0.3mm,black!60] (a4) -- (a5);
        \node [whitebox, above right = -0.6cm and 0cm of a4, minimum size = 1em, name = a7_a8]{\textbf{FPGA}};
        \node [whitebox, below right = -0.6cm and 0cm of a4, minimum size = 1em, name = a7_a8]{\textbf{Bitstream}};



    \end{tikzpicture}
}

\newcommand{\toolarchnew}[0]{
    \begin{tikzpicture}[auto, node distance = 1cm, scale = 1, transform shape, font = \bf \scriptsize \sffamily]

        \node[fancynodesqr, name = distributor, font = \normalsize ]{\textbf{D}};

        \node[fancycircle, above right =  0.8cm and 1cm of distributor, name = u1]{\textbf{U1}};
        \node[fancycircle, above right = -0.2cm and 1cm of distributor, name = u2]{\textbf{U2}};
        \node[fancycircle, below right = -0.2cm and 1cm of distributor, name = u3]{\textbf{U3}};
        \node[fancycircle, below right =  0.8cm and 1cm of distributor, name = u4]{\textbf{U4}};

        \node[fancynodelong, right = 0.5cm of u1, name = a1]{\textbf{A1}};
        \node[fancynodelong, right = 0.5cm of u2, name = a2]{\textbf{A2}};
        \node[fancynodelong, right = 0.5cm of u3, name = a3]{\textbf{A3}};
        \node[fancynodelong, right = 0.5cm of u4, name = a4]{\textbf{A4}};

        \node[fancycircle, right = 0.5cm of a1, name = p1]{\textbf{P1}};
        \node[fancycircle, right = 0.5cm of a2, name = p2]{\textbf{P2}};
        \node[fancycircle, right = 0.5cm of a3, name = p3]{\textbf{P3}};
        \node[fancycircle, right = 0.5cm of a4, name = p4]{\textbf{P4}};

        \node[fancynodesqr, right = 5.7cm of distributor, name = collector, font = \normalsize]{\textbf{C}};

        \node[group, fit=(u1)(a1)(p1)](g1){};
        \node[group, fit=(u2)(a2)(p2)](g2){};
        \node[group, fit=(u3)(a3)(p3)](g3){};
        \node[group, fit=(u4)(a4)(p4)](g4){};

        \node[whitebox, right = 2.3cm of distributor, name = ellipsis1] {\LARGE{. . .}};

        \draw [->, rounded corners, thick, line width=0.4mm] (distributor)
            to[out=0,  in=180] ++(1.0cm, 0.0cm)
            to[out=90, in=-90] ++(0.0cm, 1.75cm) to (g1);
        \draw [->, rounded corners, thick, line width=0.4mm] (distributor)
            to[out=0,  in=180] ++(1.0cm, 0.0cm)
            to[out=90, in=-90] ++(0.0cm, 0.75cm) to (g2);
        \draw [->, rounded corners, thick, line width=0.4mm] (distributor)
            to[out=0,  in=180] ++(1.0cm, 0.0cm)
            to[out=-90, in=90] ++(0.0cm, -0.75cm) to (g3);
        \draw [->, rounded corners, thick, line width=0.4mm] (distributor)
            to[out=0,  in=180] ++(1.0cm, 0.0cm)
            to[out=-90, in=90] ++(0.0cm, -1.75cm) to (g4);

        \draw [->, line width = 0.4mm] (u1) -- (a1);
        \draw [->, line width = 0.4mm] (u2) -- (a2);
        \draw [->, line width = 0.4mm] (u3) -- (a3);
        \draw [->, line width = 0.4mm] (u4) -- (a4);

        \draw [->, line width = 0.4mm] (a1) -- (p1);
        \draw [->, line width = 0.4mm] (a2) -- (p2);
        \draw [->, line width = 0.4mm] (a3) -- (p3);
        \draw [->, line width = 0.4mm] (a4) -- (p4);

        \draw [<-, rounded corners, thick, line width=0.4mm] (collector)
            to[out=180, in=0] ++(-1.0cm, 0.0cm)
            to[out=90, in=-90] ++(0.0cm, 1.75cm) to (g1);
        \draw [<-, rounded corners, thick, line width=0.4mm] (collector)
            to[out=180,  in=0] ++(-1.0cm, 0.0cm)
            to[out=90, in=-90] ++(0.0cm, 0.75cm) to (g2);
        \draw [<-, rounded corners, thick, line width=0.4mm] (collector)
            to[out=180,  in=0] ++(-1.0cm, 0.0cm)
            to[out=-90, in=90] ++(0.0cm, -0.75cm) to (g3);
        \draw [<-, rounded corners, thick, line width=0.4mm] (collector)
            to[out=180,  in=0] ++(-1.0cm, 0.0cm)
            to[out=-90, in=90] ++(0.0cm, -1.75cm) to (g4);

        \node[fancynodesqr, below left = 1.9cm and -0.6cm of distributor, name = D, font = \scriptsize, minimum height = 1.5em, minimum width = 1.5em]{\textbf{D}};
        \node[whitebox, right = 0.1cm of D, name = Ddef] {Distributor};

        \node[fancynodesqr, below = 0.1cm of D, name = C, font = \scriptsize, minimum height = 1.5em, minimum width = 1.5em]{\textbf{C}};
        \node[whitebox, right = 0.1cm of C, name = Cdef] {Collector};

        \node[fancycircle, right = 2cm of D, minimum size = 0.8em, name = U]{\textbf{U}};
        \node[whitebox, right = 0.1cm of U, name = Udef] {Unpack};

        \node[fancycircle, below = 0.08cm of U, minimum size = 0.8em, name = P]{\textbf{P}};
        \node[whitebox, right = 0.1cm of P, name = Pdef] {Pack};

        \node[fancynodelong, right = 4.3cm of D, minimum width = 2.4em, minimum height = 1.2em, name = A]{\textbf{A}};
        \node[whitebox, right = 0.1cm of A, name = Adef] {Accelerator};

        \node[group, below = 0.2cm of A, minimum width = 2.4em, minimum height = 1.2em, name = G] {};
        \node[whitebox, right = 0.1cm of G, name = Gdef] {Processing Element};

    \end{tikzpicture}
}

\newcommand{\toolarch}[0]{
    \begin{tikzpicture}[node distance = 3cm, scale=0.43,transform shape]
    \node at (0,0) (input) {};
    \node[block, right of=input](distributor) {distributor};
    \draw[->,>=stealth'] (input) -- (distributor) node[midway,above] {VL};

    \foreach \i in {1,2}
    {
        \ifthenelse{\i=1}
        {\node[block, above right = of distributor](splitter\i) {unpack};
        \node[block, right of = splitter\i] (acc\i) {acc\i};}
        {\node[block, below right = of distributor](splitter\i) {unpack};
        \node[block, right of = splitter\i] (acc\i) {accP};}
        \node[block, right of = acc\i] (assembler\i) {pack};
        \draw[->,>=stealth'] (splitter\i) to (acc\i);
        \draw[->,>=stealth'] (acc\i) to (assembler\i);
        \draw[->,>=stealth'] (distributor.east) to (splitter\i.west);
        \ifthenelse{\i=1}
        {\draw[dashed, color=orange!50] ([xshift = -0.75cm,yshift = -0.75cm]splitter\i.south west) rectangle ([xshift = 0.75cm,yshift = 0.75cm]assembler\i.north east) node[font=\fontsize{20}{15}\selectfont, left=3pt, color=black]{PE \i};}
        {\draw[dashed, color=orange!50] ([xshift = -0.75cm,yshift = -0.75cm]splitter\i.south west) rectangle ([xshift = 0.75cm,yshift = 0.75cm]assembler\i.north east) node[font=\fontsize{20}{15}\selectfont, left=3pt, color=black]{PE P};}
    }
    \node[right of = distributor](splitteri){};
    \draw[->,>=stealth'] (distributor) -- (splitteri);

    \node[block, below right = of assembler1] (collector) {collector};
    \node[right of = collector](output){};
    \draw[->,>=stealth'] (collector) -- (output) node[midway,above] {VL};
    \draw[->,>=stealth'] (assembler1.east) to (collector.west);
    \draw[->,>=stealth'] (assembler2.east) to (collector.west);
    \draw[dotted] ([yshift = -1cm]acc1.south) to ([yshift = 1cm]acc2.north);
    \end{tikzpicture}
}

\newcommand{\harrisHist}[0]{
            \begin{tikzpicture}[scale=0.6]
\begin{axis}[
	cycle list name=exotic,
	ylabel={\textbf{\% Pixels}},
	xlabel={\textbf{Bits required}},
        	ymin=0, ymax=100,
        	xmin=2, xmax=8,
        	axis y line*=left,
      	axis x line*=bottom,
      	y axis line style={opacity=0.1},
      	x axis line style={opacity=0.1},
    	ymajorgrids=true,
    	xmajorgrids=true,
       	legend style=
       		{
       		draw=none,
        	  	at={(0.7,0.1)},
         	   	anchor=south,
         	 	legend columns=1,
        	 	/tikz/every even column/.append style={column sep=0.05cm}
 	 	},
 	every tick label/.append style={font=\small},
        	xtick={2,3,4,5,6,7,8,9,10,11,12,13,14},
        	ytick={0,10,20,30,40,50,60,70,80,90,100},
]

\addplot+[line width=1.1pt] plot coordinates {
  (2, 56.14206429)
  (3, 72.70465)
  (4, 84.73822857)
  (5, 92.81682857)
  (6, 97.81227857)
  (7, 99.91657857)
  (8 ,100)
};
\addlegendentry{Stage $I_x$}

\addplot+[line width=1.1pt] plot coordinates {
  (2,  44.94556429)
  (3, 56.14206429)
  (4 ,64.29742143)
  (5 ,72.70465)
  (6 ,79.04923571)
  (7 ,84.73822857)
  (8 ,89.15143571)
  (9 ,92.81682857)
  (10,  95.65925714)
  (11 , 97.81227857)
  (12  ,99.26000714)
  (13  ,99.91657857)
  (14  ,100)
};
\addlegendentry{Stage $I_{xy}$}

\end{axis}
\end{tikzpicture}
\label{fig:hs1}
            \label{fig:hs2}
}

\newcommand{\harrisProfPower}[0]{
        \begin{tikzpicture}[scale=0.7]
        \pgfplotsset{
            cycle list name=exotic,
            scale only axis,
            xmin=0, xmax=14,
            x axis line style={opacity=0.1},
            y axis line style={opacity=0.1},
            xmajorgrids=true,
            every tick label/.append style={font=\small},
            xtick={0,2,4,6,8,10,12,14},
            legend style=
                    {
                        at={(0.8,0.05)},
                        anchor=south,
                        legend columns=1,
                        /tikz/every even column/.append style={column sep=0.05cm}
                }
        }

        \begin{axis}[
            ylabel={\textbf{Power (W)}},
            xlabel={\textbf{Number of fractional bits ($\beta$)}},
                    ymin=0.20, ymax=0.4,
                    axis y line=right,
                ymajorgrids=true,
                    ytick={0.20,0.22,0.24,0.26,0.28,0.30,0.32,0.34,0.36,0.38,0.40},
        ]

        \addplot+[thick,line width=1.1pt] coordinates {
        (0, 0.246*0.976)
        (2, 0.229*0.976)
        (4, 0.296*0.976)
        (6, 0.258*0.976)
        (8, 0.297*0.976)
        (10,    0.292*0.976)
        (12,    0.356*0.976)
        (14,    0.337*0.976)

        }; \label{plot_one_hcd}
        \end{axis}

        \begin{axis}[
                    ymin=99, ymax=100,
                    axis y line=left,
                    axis x line=none,
                    ylabel={\textbf{Correctly Classified Corners (\%)}},
                ymajorgrids=true,
        ]

        \addplot plot 
        coordinates {
            (0, 99.571)
        };
        \addlegendentry{Power}

        \addplot+[thick,line width=1.1pt] coordinates {
            
            (0, 99.571)
            (2, 99.8967)
            (4, 99.9724)
            (6, 99.9926)
            (8, 99.9979)
            (10,    99.9994)
            (12,    99.9998)
            (14,    99.9999)

        }; \label{plot_two_hcd}
        \addlegendentry{Quality}

        \addlegendimage{black} 

        \end{axis}
        \end{tikzpicture}
}

\newcommand{\mountain}[0]{
            \subfloat[]{
                \includegraphics[trim={0 0 0 8cm},clip]{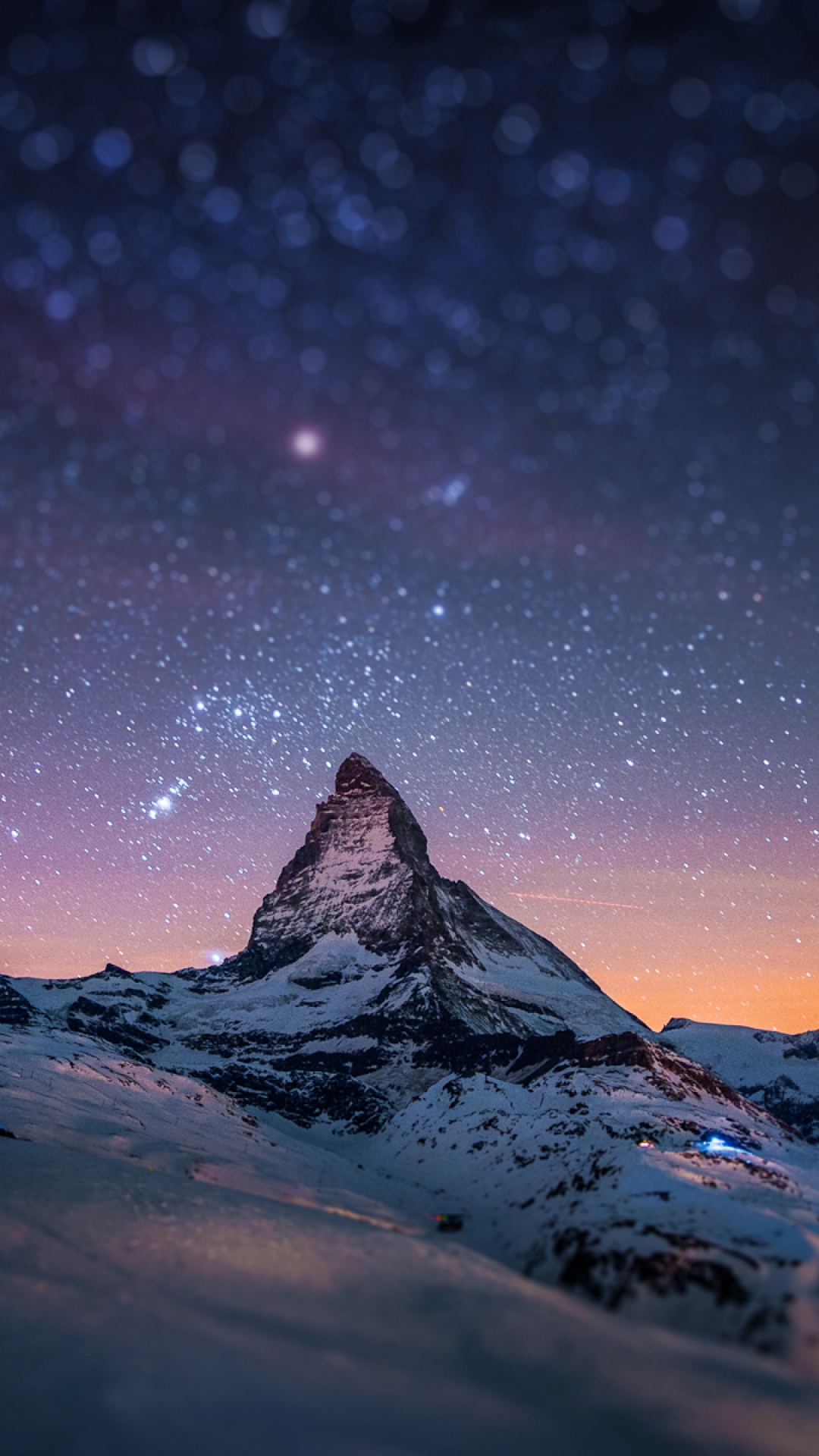}
                \label{fig:mount}
        }
            \subfloat[]{
                \includegraphics[trim={0 0 0 8cm},clip]{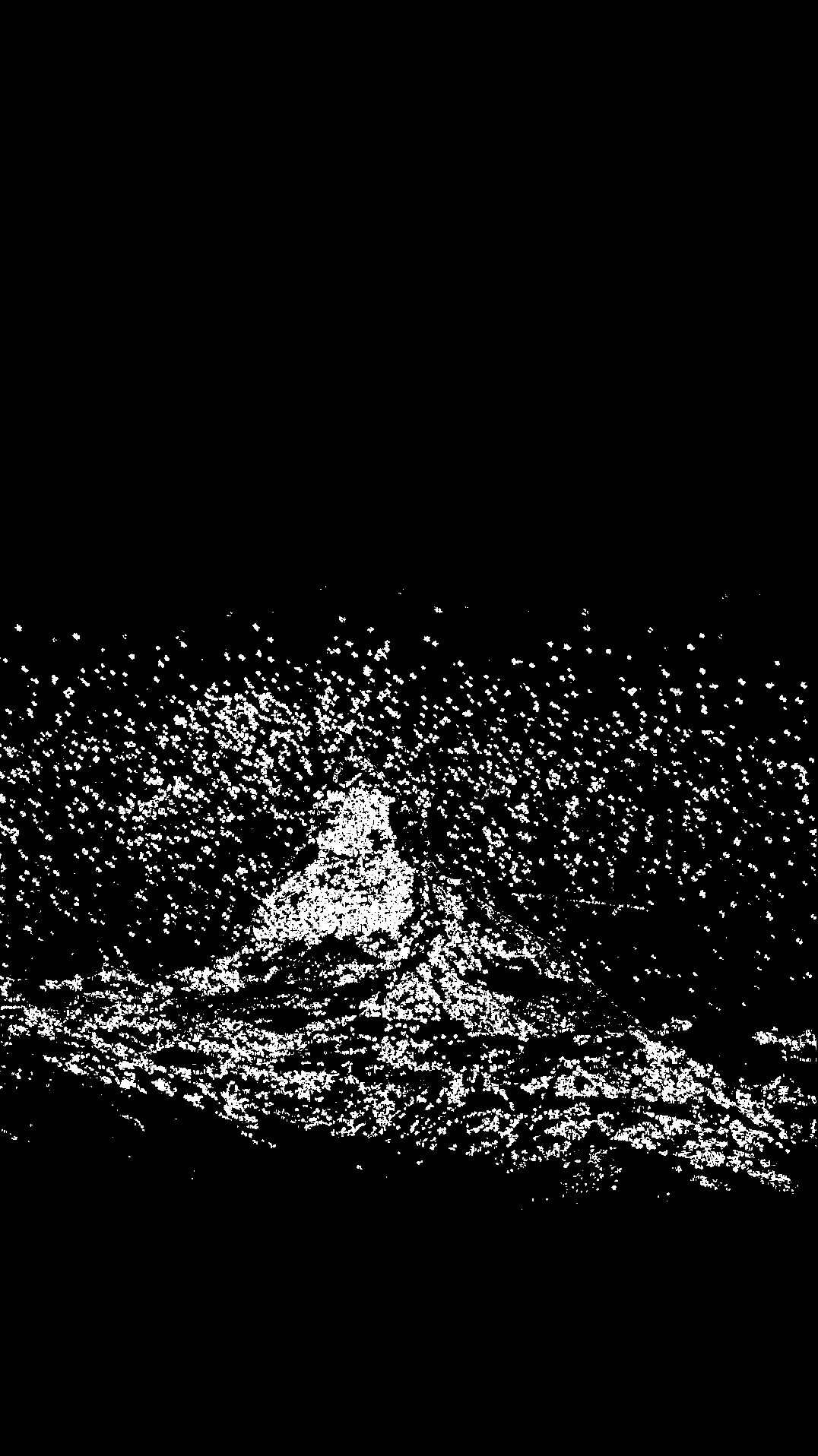}
                \label{fig:mountHcd}
        }
}
\newcolumntype{?}{!{\vrule width 0.8pt}}
\newcolumntype{$}{!{\vrule width 0.8pt}}

\newcommand{\Ress}[0]{
        \centering
\resizebox{\linewidth}{!}{
        \begin{tabularx}{1.4\linewidth}{c c c c  c c c c c c c c c c c}
            \toprule
        & Analysis & \multicolumn{2}{c}{Quality} & Power   & Clk   &  Latency  & BRAM  & DSP & FF & LUT & FPGA slices &  Min.Clk  & Max.Throughput & Power\\
                    & & (unit) & (value)         & (Watts) & (ns)  & (million) & Tiles &     &    &     & used (\%) &  (ns)  & (MPixels/sec) & (Watts)\\
        \specialrule{0.1em}{.4em}{.4em}
        O & Float  && 0.17                        & 0.641   &      6 &2.07 &             34&    168&   33366&  20520& 65.28 &                  5.84  & 171 & 0.655  \\ 
        F & $\ \ \alpha^{Z3RA}$                &AAE& 1.60          & 0.328   &      6 &2.07 &             20&     22&  11810 &   7497& 26.20 & 4.68  & 214 & 0.414  \\ 
          & $\ \ \alpha^{IA}$ &(in degrees)& 1.60 & 0.459   &      6 &2.07 &             42&     44&   20796&  11440&  46.32 &                 4.88  & 205 & 0.574  \\ 
          & $ \ \ \alpha^{avgP}$ & & 1.60          & 0.311   &      6 &2.07 &             18&     22&   11496&   7212& 26.17 &                 4.54  & 220 & 0.398  \\
        \specialrule{0.01em}{.4em}{.4em}
        H & Float  &  &99.999 & 0.970  &                5.5 & 2.06 & 32&   113&   18420&  22961& 33.01  & 5.24&190 & 0.956     \\
        C & $ \ \ \alpha^{IA}$ & \% &99.999 & 0.263 &     5.5 & 2.06 &  14&     12&    2902&   2724& 5.35  & 4.76&210    &   0.369    \\
        D & $ \ \ \alpha^{avgP}$ &  & 99.999 & 0.253 & 5.5 & 2.06 &   14 &    12&    2848&   2727& 5.32 & 4.68&214    &   0.357    \\
        \specialrule{0.01em}{.4em}{.4em}
        D & Float  &  & Inf & 0.269  &               5 & 6.22&   14&54&9150&10061& 13.54              &   4.84&206  & 0.271    \\
        U & $ \ \ \alpha^{IA}$ &PSNR& Inf & 0.159 &    5 & 6.22&    7& 0&    5161&   2744& 3.39  &4.26 &234               &  0.166   \\
        S & $ \ \ \alpha^{avgP}$  &&Inf & 0.159 &  5 & 6.22&    7& 0&    5161&   2744& 3.39  & 4.26& 234              &  0.166   \\
        \specialrule{0.01em}{.4em}{.4em}
        U & Float  &&21e-6, 0.05  &  0.273 & 4.5 & 6.22 & 8&     46&    7452&   9012& 12.38 & 4.33 & 230   & 0.288       \\
        S & $ \ \ \alpha^{IA}$ &classification,& 0.05, 0.16  &  0.169  & 4.5 & 6.22 &   4&      2&    3010&   2984& 4.71 & 4.08 & 245    &   0.175      \\
        M & $ \ \ \alpha^{avgP}$ &rms& 0.05, 0.16 & 0.169 &  4.5 & 6.22 &  4&      2&    3010&   2984& 4.71 &  4.08 &245     &   0.175     \\
        \Xhline{2.3\arrayrulewidth}
        \end{tabularx}
        }
}

\newcommand{\dusRes}[0]{
\resizebox{\linewidth}{!}{
        \begin{tabularx}{1.12\linewidth}{c c c c c c c c c c| c c c}
        \toprule
        Analysis & Quality & Power   & Clk Period   &  Latency  &BRAM & DSP & FF & LUT & Zedboard FPGA &  Min. Clk Period  & Max. Throughput & Power\\
                    & (PSNR) & (Watts)   & (ns)&  (Million)  & &&& & slices used(\%) &  (ns)  & (MPixels/sec) & (Watts)\\
        \midrule
        Float  &  Inf & 0.269  &               5 & 6.22&   14&54&9150&10061& 13.54              &   4.84&206  & 0.271    \\
        $ \ \ \alpha^{sa}$ & Inf & 0.159 &    5 & 6.22&    7& 0&    5161&   2744& 3.39  &4.26 &234               &  0.166   \\
        $ \ \ \alpha^{avg}$ & Inf & 0.159 &  5 & 6.22&    7& 0&    5161&   2744& 3.39  & 4.26& 234              &  0.166   \\
        \bottomrule
        \end{tabularx}
        }
}

\newcommand{\usmRes}[0]{
\resizebox{\linewidth}{!}{
        \begin{tabularx}{1.2\linewidth}{c c c c c c c c c c| c c c }
        \toprule
        Analysis & Quality & Power   & Clk Period   &  Latency  &BRAM & DSP & FF & LUT & Zedboard FPGA &  Min. Clk Period  & Max. Throughput & Power\\
                    & (classification,rms) & (Watts) & (ns)&  (Million)  & &&& & slices used(\%) &  (ns)  & (MPixels/sec) & (Watts)\\
        \midrule
        Float  &21e-6 , 0.05  &  0.273 & 4.5 & 6.22 & 8&     46&    7452&   9012& 12.38 & 4.33 & 230   & 0.288       \\
        $ \ \ \alpha^{sa}$ & 0.05 , 0.16  &  0.169  & 4.5 & 6.22 &   4&      2&    3010&   2984& 4.71 & 4.08 & 245    &   0.175      \\
        $ \ \ \alpha^{avg}$ & 0.05 , 0.16 & 0.169 &  4.5 & 6.22 &  4&      2&    3010&   2984& 4.71 &  4.08 &245     &   0.175     \\
        \bottomrule
        \end{tabularx}
        }
}

\newcommand{\ofRes}[0]{
\resizebox{\linewidth}{!}{
        \begin{tabularx}{1.2\linewidth}{c c c c c c c c c c| c c c}
        \toprule
        Analysis & Quality & Power   & Clk Period   &  Latency  &BRAM & DSP & FF & LUT & Zedboard FPGA &  Min. Clk Period  & Max. Throughput & Power\\
                    & (AAE (in degree)) & (Watts) & (ns)&  (Million)  & &&& & slices used(\%) &  (ns)  & (MPixels/sec) & (Watts)\\
        \midrule
        Float  & 0.17 & 0.641  &                   6 &2.07 &             68& 168&   44196&  43548& 65.28   &   5.54  &   179  &  0.683    \\ 
        $\ \ \alpha^{sa}$ & 1.60 & 0.398   &      6 &2.07 &             60&     68&   35322&  22013& 32.33 &   4.90  &   203  &    0.570   \\ 
        $ \ \ \alpha^{avg}$ & 1.60 & 0.388 &     6 &2.07 &           36& 22&   27495&  18670&  25.69      &   4.63  &   215  &    0.469   \\ 
        \bottomrule
        \end{tabularx}
        }
}

\newcommand{\dusDag}[0]{
    \begin{tikzpicture}
        \tikzset{VertexStyle/.append style = {
                minimum size = 2cm,font = \fontsize{17}{17}\bfseries},
                EdgeStyle/.append style = {->,>=stealth'},
                LabelStyle/.style = {above=2pt, text = blue!20!black, font = \fontsize{12}{12}\bfseries }}
        \SetGraphUnit{3.5}
        \Vertices[Math]{line}{img,D_x,D_y,U_x,U_y}

        \Edge(img)(D_x)
        \Edge(D_x)(D_y)
        \Edge(D_y)(U_x)
        \Edge(U_x)(U_y)
    \end{tikzpicture}
}

\newcommand{\dusMatrixb}[0]{
\centering
    $\begin{bmatrix}
        1 & 3 & 3 & 1 \\
    \end{bmatrix} / 8 $
}

\newcommand{\dusMatrixc}[0]{
    $\begin{bmatrix}
    1&3\\
    \end{bmatrix} / 4 $
}

\newcommand{\usmDag}[0]{
    \begin{tikzpicture}
        \tikzset{VertexStyle/.append style = {
                minimum size = 2cm,font = \fontsize{17}{17}\bfseries},
                EdgeStyle/.append style = {->,>=stealth'},
                LabelStyle/.style = {above=2pt, text = blue!20!black, font = \fontsize{12}{12}\bfseries }}
        \SetGraphUnit{3.5}
        \Vertices[Math]{line}{img,blurx,blury,sharpen,mask}

        \Edge[](img)(blurx)
        \Edge[style={bend left}](img)(sharpen)
        \Edge[style={bend left}](img)(mask)
        \Edge[](blurx)(blury)
        \Edge[](blury)(sharpen)
        \Edge[style={bend right}](blury)(mask)
        \Edge[](sharpen)(mask)
        \tikzset{LabelStyle/.append style = {below=2pt}}
    \end{tikzpicture}
}

\newcommand{\pipeunsharp}[0]{
    \begin{tikzpicture}[auto, node distance=1.25cm,>=latex', scale=1, transform shape]
        \node [input, name=input] {\textbf{img}};
        \node [fancynodelong, below of=input] (blurx) {\textbf{blurx}};
        \node [fancynodelong, below of=blurx] (blury) {\textbf{blury}};
        \node [fancynodelong, below of=blury] (sharpen) {\textbf{sharpen}};
        \node [fancynodelong, below of=sharpen] (masked) {\textbf{mask}};

        \node [above left = -2.7cm and 1.85cm of input, name=inpic] {
            \includegraphics[width=3.5cm]{images/image.jpg}
        };

        \node [below = 0.1cm of inpic, name=outpic] {
            \includegraphics[width=3.5cm]{images/image_sharpened.jpg}
        };

        \draw [<-, thickflowarrow] (input) to [out=180, in=0] ++(-2.3, 0);
        \draw [->, flowarrow] (input) -- (blurx);
        \draw [->, flowarrow] (input.east) to [bend left =45] (sharpen);
        \draw [->, flowarrow] (input.east) to [bend left =45] (masked);
        \draw [->, flowarrow] (blurx) -- (blury);
        \draw [->, flowarrow] (blury) -- (sharpen);
        \draw [->, flowarrow] (blury.west) to [bend right=45] (masked);
        \draw [->, flowarrow] (sharpen) -- (masked);
        \draw [->, thickflowarrow] (masked) to [out=180, in=0] ++(-2.35, 0);

    \end{tikzpicture}
}

\newcommand{\usmMatrixb}[0]{
    $\begin{bmatrix}
        1 & 4 & 6 & 4 & 1 \\
    \end{bmatrix} / 16 $
}

\newcommand{\usmProfPower}[0]{
        \begin{tikzpicture}[scale=0.5]
        \pgfplotsset{
            cycle list name=exotic,
            scale only axis,
            xmin=0, xmax=14,
            x axis line style={opacity=0.1},
            y axis line style={opacity=0.1},
            xmajorgrids=true,
            every tick label/.append style={font=\small},
            xtick={0,2,4,6,8,10,12,14},
            legend style=
                    {
                        at={(0.7,0.14)},
                        anchor=south,
                        legend columns=1,
                        /tikz/every even column/.append style={column sep=0.05cm}
                }
        }

        \begin{axis}[
            ylabel={\textbf{Power (W)}},
            xlabel={\textbf{Number of fractional bits ($\beta$)}},
                    ymin=0.0, ymax=0.19,
                    axis y line=right,
                ymajorgrids=true,
        ]

        \addplot+[thick,line width=1.1pt] coordinates {
            (0,   0.164)
            (1,   0.173)
            (2,   0.173)
            (3,   0.174)
            (4,   0.169)
            (5,   0.176)
            (6,   0.178)
            (7,   0.177)
            (8,   0.173)
            (9,   0.174)
            (10,  0.174)
            (11,  0.184)
            (12,  0.185)
            (13,  0.183)
            (14,  0.183)
        }; \label{plot_one_usm}
        \end{axis}

        \begin{axis}[
                    ymin=0, ymax=0.8,
                    axis y line=left,
                    axis x line=none,
                    ylabel={\textbf{Error}},
                ymajorgrids=true,
        ]

        \addplot plot 
        coordinates {
            (0,    14.19713946/100)
        };
        \addlegendentry{Power}

        \addplot+[thick,line width=1.1pt] coordinates {
            
            (0, 14.19713946/100)
        (1 ,  9.972670068/100)
        (2  , 6.584627551/100)
        (3   ,4.307321429/100)
        (4   ,1.313392857/100)
        (5,    0.6779676871/100)
        (6 ,   0.2759268707/100)
        (7  ,  0.1324880952/100)
        (8   , 0.0003911476179/100)
        (9 ,   0.0002567938714/100)
        (10,   0.0001292442179/100)
        (11 ,  0.0001292442179/100)
        (12,   0.0001173428571/100)
        (13,   0.00009523468929/100)
        (14,   0.00005952040714/100)

        }; \label{plot_two_usm}
        \addlegendentry{Misclassified bits fraction}

        \addplot[thick,cyan!50!blue,every mark/.append style={fill=cyan!80!black},mark=diamond*,line width=1.1pt,mark size=2.5pt] coordinates {
            
         (0,  0.5961569979)
         (1,  0.3974640616)
         (2,  0.3027937173)
         (3,  0.2132470336)
         (4,  0.1657459729)
         (5,  0.1070057421)
         (6,  0.07635584998)
         (7,  0.0452333177)
         (8 , 0.0258403306)
         (9,  0.0199785589)
         (10, 0.01167980521)
         (11, 0.01165446643)
         (12, 0.0114356684)
         (13, 0.01076936252)
         (14, 0.009698866118)

        }; \label{plot_three_usm}
        \addlegendentry{RMS error}

        \addlegendimage{black} 

        \end{axis}
        \end{tikzpicture}
}

\newcommand{\dusProfPower}[0]{
        \begin{tikzpicture}[scale=0.5]
        \pgfplotsset{
            cycle list name=exotic,
            scale only axis,
            xmin=0, xmax=12,
            x axis line style={opacity=0.1},
            y axis line style={opacity=0.1},
            xmajorgrids=true,
            every tick label/.append style={font=\small},
            xtick={0,2,4,6,8,10,12},
            legend style=
                    {
                        at={(0.8,0.05)},
                        anchor=south,
                        legend columns=1,
                        /tikz/every even column/.append style={column sep=0.05cm}
                }
        }

        \begin{axis}[
            ylabel={\textbf{Power (W)}},
            xlabel={\textbf{Number of fractional bits ($\beta$)}},
                    ymin=0.1, ymax=0.20,
                    axis y line=right,
                ymajorgrids=true,
        ]

        \addplot+[thick,line width=1.1pt] coordinates {
       
       (0,    0.156*0.93)
       (1 ,   0.155*0.93)
       (2  ,  0.158*0.93)
       (3   , 0.163*0.93)
       (4 ,   0.163*0.93)
       (5  ,  0.164*0.93)
       (6   , 0.164*0.93)
       (7 ,   0.167*0.93)
       (8  ,  0.166*0.93)
       (9   , 0.167*0.93)
       (10,   0.171*0.93)
       (11 ,  0.189*0.93)
       (12  , 0.184*0.93)

        }; \label{plot_one_dus}
        \end{axis}

        \begin{axis}[
                    ymin=40, ymax=100,
                    axis y line=left,
                    axis x line=none,
                    ylabel={\textbf{Quality}},
                    ymajorgrids=true,
                    ytick= {40,50,60,70,80,90,100},
                    yticklabels={40,50,60,70,80,90,\large{$\infty$}},
        ]

        \addplot plot 
        coordinates {
            (0, 44.9105)
        };
        \addlegendentry{Power}

        \addplot+[thick,line width=1.1pt] coordinates {
            
           (0,    44.91053262)
           (1,    46.15428085)
           (2,    46.86602746)
           (3,    48.15499744)
           (4,    49.19576655)
           (5,    50.37276094)
           (6,    53.17879168)
           (7,    55.17239188)
           (8,    59.80610038)
           (9,    62.62903722)
           (10,   100)
           (11,   100)
           (12,   100)

        }; \label{plot_two_dus}
        \addlegendentry{Quality (PSNR)}

        \addlegendimage{black} 

        \end{axis}
        \end{tikzpicture}
}

\newcommand{\ofProfPower}[0]{
        \begin{tikzpicture}[scale=0.5]
        \pgfplotsset{
            cycle list name=exotic,
            scale only axis,
            xmin=4, xmax=26,
            x axis line style={opacity=0.1},
            y axis line style={opacity=0.1},
            xmajorgrids=true,
            every tick label/.append style={font=\small},
            xtick= {2,4,6,8,10,14,18,22,26},
            legend style=
                    {
                        at={(0.7,0.14)},
                        anchor=south,
                        legend columns=1,
                        /tikz/every even column/.append style={column sep=0.05cm}
                }
        }

        \begin{axis}[
            ylabel={\textbf{Power (W)}},
            xlabel={\textbf{Number of fractional bits ($\beta$)}},
                    ymin=0, ymax=1.0,
                    axis y line=right,
                ymajorgrids=true,
        ]

        \addplot+[thick,line width=1.1pt] coordinates {
            (4,   0.322*0.946)
            (8,  0.351*0.946)
            (10,  0.41*0.946)
            (14,  0.545*0.946)
            (18,  0.627*0.946)
            (22,  0.719*0.946)
            (26,  0.878*0.946)
        }; \label{plot_one_of}
        \end{axis}

        \begin{axis}[
                    ymin=0, ymax=12,
                    axis y line=left,
                    axis x line=none,
                    ylabel={\textbf{Error}},
                ymajorgrids=true,
        ]

        \addplot plot 
        coordinates {
            (4,    0.1851255)
        };
        \addlegendentry{Power}

        \addplot+[thick,line width=1.1pt] coordinates {
            
        (4,   0.1851255114*57.2958)
        (6 ,  0.05824303426*57.2958)
        (8  , 0.03476971546*57.2958)
        (10,  0.02994369546*57.2958)
        (14,  0.02825591115*57.2958)
        (18,  0.02857969176*57.2958)
        (22,  0.02871823056*57.2958)
        (26,  0.02863478889*57.2958)

        }; \label{plot_two_of}
        \addlegendentry{AAE (in degree)}

        \addlegendimage{black} 

        \end{axis}
        \end{tikzpicture}
}

\newcommand{\usmHist}[0]{
            \subfloat[Stage $blur_x$]{\begin{tikzpicture}[scale=0.70]
\begin{axis}[
	cycle list name=exotic,
	ylabel={\textbf{\% Pixels}},
	xlabel={\textbf{Bits occupied}},
        	ymin=0, ymax=100,
        	xmin=2, xmax=8,
        	axis y line*=left,
      	axis x line*=bottom,
      	y axis line style={opacity=0.1},
      	x axis line style={opacity=0.1},
    	ymajorgrids=true,
    	xmajorgrids=true,
       	legend style=
       		{
       		draw=none,
        	  	at={(0.7,0.1)},
         	   	anchor=south,
         	 	legend columns=1,
        	 	/tikz/every even column/.append style={column sep=0.05cm}
 	 	},
 	every tick label/.append style={font=\small},
        	xtick={2,3,4,5,6,7,8},
        	ytick={0,10,20,30,40,50,60,70,80,90,100},
]

\addplot+[line width=1.1pt] plot coordinates {
  (1, 0.323)
  (2, 1.97)
  (3, 4.642989043)
  (4, 9.735566842)
  (5, 19.70236915)
  (6, 34.42375923)
  (7, 64.57803077)
  (8, 100)
};
\end{axis}
\end{tikzpicture}}
            \label{fig:usm1}
            \subfloat[Stage $blur_y$]{\begin{tikzpicture}[scale=0.70]
\begin{axis}[
	cycle list name=exotic,
	ylabel={\textbf{\% Pixels}},
	xlabel={\textbf{Bits occupied}},
        	ymin=0, ymax=100,
        	xmin=2, xmax=8,
        	axis y line*=left,
      	axis x line*=bottom,
      	y axis line style={opacity=0.1},
      	x axis line style={opacity=0.1},
    	ymajorgrids=true,
    	xmajorgrids=true,
       	legend style=
       		{
       		draw=none,
        	  	at={(0.7,0.1)},
         	   	anchor=south,
         	 	legend columns=1,
        	 	/tikz/every even column/.append style={column sep=0.05cm}
 	 	},
 	every tick label/.append style={font=\small},
        	xtick={2,3,4,5,6,7,8},
        	ytick={0,10,20,30,40,50,60,70,80,90,100},
]

\addplot+[line width=1.1pt] plot coordinates {
  (1, 0.2816633792)
  (2, 1.919337814)
  (3, 4.418082335)
  (4, 9.421442615)
  (5, 19.20815908)
  (6, 33.95212231)
  (7, 64.77952308)
  (8, 100)
};
\end{axis}
\end{tikzpicture}
            \label{fig:usm2}
            } \\ 
            \subfloat[Stage $sharpen$]{\begin{tikzpicture}[scale=0.70]
\begin{axis}[
	cycle list name=exotic,
	ylabel={\textbf{\% Pixels}},
	xlabel={\textbf{Bits occupied}},
        	ymin=0, ymax=100,
        	xmin=2, xmax=9,
        	axis y line*=left,
      	axis x line*=bottom,
      	y axis line style={opacity=0.1},
      	x axis line style={opacity=0.1},
    	ymajorgrids=true,
    	xmajorgrids=true,
       	legend style=
       		{
       		draw=none,
        	  	at={(0.7,0.1)},
         	   	anchor=south,
         	 	legend columns=1,
        	 	/tikz/every even column/.append style={column sep=0.05cm}
 	 	},
 	every tick label/.append style={font=\small},
        	xtick={2,3,4,5,6,7,8,9},
        	ytick={0,10,20,30,40,50,60,70,80,90,100},
]

\addplot+[line width=1.1pt] plot coordinates {
  (1, 0.79755682)
  (2, 2.388816992)
  (3, 5.279017269)
  (4, 10.49035846)
  (5, 20.66855385)
  (6, 35.26189308)
  (7, 64.28426154)
  (8, 99.55982308)
  (9, 100)
};
\end{axis}
\end{tikzpicture}
            \label{fig:usm3}
            }
            \subfloat[Stage $mask$]{\begin{tikzpicture}[scale=0.70]
\begin{axis}[
	cycle list name=exotic,
	ylabel={\textbf{\% Pixels}},
	xlabel={\textbf{Bits occupied}},
        	ymin=0, ymax=100,
        	xmin=2, xmax=8,
        	axis y line*=left,
      	axis x line*=bottom,
      	y axis line style={opacity=0.1},
      	x axis line style={opacity=0.1},
    	ymajorgrids=true,
    	xmajorgrids=true,
       	legend style=
       		{
       		draw=none,
        	  	at={(0.7,0.1)},
         	   	anchor=south,
         	 	legend columns=1,
        	 	/tikz/every even column/.append style={column sep=0.05cm}
 	 	},
 	every tick label/.append style={font=\small},
        	xtick={2,3,4,5,6,7,8},
        	ytick={0,10,20,30,40,50,60,70,80,90,100},
]

\addplot+[line width=1.1pt] plot coordinates {
  (1, 49.63727692)
  (2, 67.65052308)
  (3, 81.81184615)
  (4, 91.15916154)
  (5, 96.97650769)
  (6, 99.62434615)
  (7, 99.99834615)
  (8, 100)
};
\end{axis}
\end{tikzpicture}
            \label{fig:usm4}
            }
}

\newcommand{\ofsmtcode}[0]{
\lstdefinelanguage{Python}{
numbers=left,
numberstyle=\footnotesize,
numbersep=1em,
xleftmargin=1em,
framextopmargin=2em,
framexbottommargin=2em,
showspaces=false,
showtabs=false,
showstringspaces=false,
frame=l,
tabsize=4,
basicstyle=\ttfamily\small\setstretch{1},
commentstyle=\color{Comments}\slshape,
stringstyle=\color{Strings},
morecomment=[s][\color{Strings}]{"""}{"""},
morecomment=[s][\color{Strings}]{'''}{'''},
keywordstyle={\color{Keywords}\bfseries},
morekeywords={[2]@invariant,pylab,numpy,np,scipy},
keywordstyle={[2]\color{Decorators}\slshape},
emph={self},
emphstyle={\color{self}\slshape},
caption={Z3py code generated for Optical Flow},
label={list:of_z3},
}
\linespread{1.3}
\begin{adjustbox}{max width=\textwidth}
    \lstinputlisting[language=Python]{files/data.py}
\end{adjustbox}
}

\newcommand{\pwisesmtcode}[0]{
\lstdefinelanguage{Python}{
numbers=left,
numberstyle=\footnotesize,
numbersep=1em,
xleftmargin=1em,
framextopmargin=2em,
framexbottommargin=2em,
showspaces=false,
showtabs=false,
showstringspaces=false,
frame=l,
tabsize=4,
basicstyle=\ttfamily\small\setstretch{1},
commentstyle=\color{Comments}\slshape,
stringstyle=\color{Strings},
morecomment=[s][\color{Strings}]{"""}{"""},
morecomment=[s][\color{Strings}]{'''}{'''},
keywordstyle={\color{Keywords}\bfseries},
morekeywords={[2]@invariant,pylab,numpy,np,scipy},
keywordstyle={[2]\color{Decorators}\slshape},
emph={self},
emphstyle={\color{self}\slshape},
caption={Pseudo Code for Pointwise stage function definition},
label={list:pwise_z3},
}
\linespread{1.3}
\begin{adjustbox}{max width=\textwidth}
    \lstinputlisting[language=Python]{files/pwise.py}
\end{adjustbox}
}

\newcommand{\convsmtcode}[0]{
\lstdefinelanguage{Python}{
numbers=left,
numberstyle=\footnotesize,
numbersep=1em,
xleftmargin=1em,
framextopmargin=2em,
framexbottommargin=2em,
showspaces=false,
showtabs=false,
showstringspaces=false,
frame=l,
tabsize=4,
basicstyle=\ttfamily\small\setstretch{1},
commentstyle=\color{Comments}\slshape,
stringstyle=\color{Strings},
morecomment=[s][\color{Strings}]{"""}{"""},
morecomment=[s][\color{Strings}]{'''}{'''},
keywordstyle={\color{Keywords}\bfseries},
morekeywords={[2]@invariant,pylab,numpy,np,scipy},
keywordstyle={[2]\color{Decorators}\slshape},
emph={self},
emphstyle={\color{self}\slshape},
caption={Pseudo Code for Stencil stage function definition},
label={list:conv_z3},
}
\linespread{1.3}
\begin{adjustbox}{max width=\textwidth}
    \lstinputlisting[language=Python]{files/conv.py}
\end{adjustbox}
}

\section{Introduction}
Field-Programmable Gate Arrays (FPGAs) are suitable for accelerating
computations from several domains such as image processing, computer vision, 
and digital signal processing. When a lower or customized precision is 
desired,  FPGAs are often expected to perform better than accelerators such 
as GPUs with respect to performance delivered per unit of energy consumed.  
When an image processing pipeline such
as Harris Corner Detection (HCD) (cf. Figure~\ref{fig:hdag} and Table~\ref{tab:hcd}),
is implemented on a CPU or a GPU, a programmer is bound to choose a pre-defined
data type such as {\tt float}, {\tt int}, or {\tt short} owing to the underlying
architectural constraints. In order to avoid arithmetic overflows, the
data types have to be chosen conservatively through over-estimation. This 
leads to a wastage of memory, and hence memory bandwidth, at all levels of 
the hierarchy; furthermore, additional energy is consumed both due to data 
transfer and the higher precision in which the arithmetic is performed.  On 
the other hand, on FPGAs, it is possible to use variable length
fixed-point data types to represent the data produced and consumed at various
stages of an image processing pipeline. This saves chip area and the power
consumed by the hardware design due to the reduced precision
and internal routing logic. The other natural outcome is a better 
utilization of available on-chip memory resources.

Although, FPGAs fare extremely well on the performance per watt metric, 
their programmability has been a major hindrance in adoption. The reliance 
on hardware description languages (HDLs) such as Verilog and VHDL makes it 
extremely cumbersome for a wider programmer audience. High-Level Synthesis 
(HLS) tools, which map C, C++ code into equivalent hardware designs by 
generating HDL code automatically, have thus gained significant attention in 
the past decade. The quality of designs generated by a HLS compiler often 
depend on the analysis techniques employed. Many times programmers put in 
substantial effort to drive a HLS tool to generate a hardware design of 
their choice using suitable pragma annotations or code rewriting. There have 
been efforts to further raise the level of abstraction --- from using 
imperative languages such as C, C++ to domain-specific languages (DSLs) --- 
giving rise to the term, ultra high-level synthesis; Bacon et 
al.~\cite{bacon13acm} provide a comprehensive survey.  DSLs not only
improve programmer productivity but also exploit the richer information  
from the underlying algorithm. This facilitates compilers to generate better 
code or hardware designs using relatively simple program analysis 
techniques.

\subsection{DSLs for Image Processing Pipelines} An image processing 
pipeline can be viewed as a directed acyclic graph (DAG) of computational 
stages. Each stage transforms an
input image form into an output image form to be consumed by the subsequent
stages in the pipeline. The class of computations at each stage are simple data
parallel operations applied on all image pixels such as point-wise and stencil
computations.  Figure~\ref{fig:hdag} shows the computational DAG associated with
the Harris Corner Detection (HCD) benchmark. As the name indicates, HCD is a corner
detection algorithm, commonly used in computer vision space.
Table~\ref{tab:hcd} shows the computations at each stage of the DAG.
The source code for HCD benchmark in PolyMage DSL, whose compiler infrastructure we use
in this paper,
can be obtained at the PolyMage  GitHub repository~\cite{polymagebench}.  
If the computations in HCD  are expressed in C/C++,
then there will be a two dimensional loop associated with each stage of the DAG.
Further, these loops occur in some topologically sorted order of DAG nodes.
Thus, the rich structure in the application gets lost in the resulting C/C++
code. For example, it is hard to infer that pixels output from one stage can be
streamed to the following stage and the following stage can start computations
once it receives enough number of pixels. This observation leads to an extremely
efficient pipelined hardware architecture for the whole computational DAG and is
exploited in PolyMage-HLS compiler as in other DSL compilers for FPGAs. In this
paper, we use the fact that it is easy to infer the computations on pixels at
each stage of the DAG and further these computations are applied homogeneously on
all the pixels at the corresponding stage, to arrive at efficient range
analysis algorithms based on interval arithmetic and Satisfiability Modulo 
Theory (SMT) solvers.  It would be almost impossible to do
this if the design is expressed directly in Verilog/VHDL; and probably require
complex program analysis if we have to to achieve this on C/C++ programs as is
the case in HLS frameworks.

\subsection{Problem Description and Contributions}
The data at each stage of an image processing pipeline is represented using
a parametric fixed-point data type $\left(\alpha, \beta\right)$, where $\alpha$ 
and $\beta$
denote the number of bits used to represent integral and fractional parts.  The
objective is to minimize $\alpha$  and $\beta$ at each stage while maintaining
an application specific quality metric. The optimal value of $\alpha$ at a stage
depends on the range of values produced; whereas the optimal value of $\beta$
depends on how precision impacts the quality metric.

Range analysis which is required for estimating the integral bitwidth requirement
is a well studied problem in literature. There are several works
based on variants of interval and affine arithmetic~\cite{cong09fccm, vakili13tcad, 
zhang10jsip, stephenson00sigplan, lee06tcad, mahlke01tcad}.
The benchmarks considered in these works, such as FIR filter, Discrete Cosine
Transform, Polynomial Evaluation etc.  are mainly from the signal processing domain 
and their code size and complexity is small. These techniques are not
easily adaptable for image processing pipelines  when expressed in Verilog/VHDL,
HLS C/C++ etc. due to the large number of pixel signals present at each stage
and multiple such stages in an application. 

The first main contribution of our paper is an interval arithmetic based range analysis 
technique in the DSL compiler which exploits the fact that the computations on all
the pixels at a given stage is homogeneous to do a combined range analysis. The
second main contribution of our paper is a range analysis framework in the DSL compiler 
wherein any interval and affine arithmetic-like analysis can be incorporated with ease.

Apart from interval arithmetic based approaches, which cannot handle certain
kind of operations like divisions, techniques using powerful SMT solvers have
been proposed in the literature~\cite{ kinsman10}. However, they are applied on
extremely small benchmarks involving 2 to 3 equations. The third main
contribution of this paper is that, we  propose a range analysis technique using
SAT solvers and apply it on large benchmarks such as Optical Flow involving few
tens of DAG stages not to speak of large number of pixel signals and complex
computations in the DAG structure. This is primarily possible because our
analysis is based on the DSL specification of the benchmark as against a HDL or
C/C++ specification. Further, we show that in iterative algorithms such as
Optical Flow, conservative estimates of interval analysis will have a
debilitating effect with the increase in the iterations making them unusable in
practice; and we have to resort to SMT solvers to get accurate range estimates.

The fourth main contribution of this paper is a simple greedy heuristic search
technique for precision analysis to determine the number of bits required for
representing fractional bits at each stage of the pipeline. Finally, we present
a thorough experimental study comparing the effectiveness of interval, SMT
solver and profile guided approaches with respect to power, area and speed on
large image processing benchmarks. We would like to highlight that all the
previous studies involve very small benchmarks. 

We implement and evaluate our automatic bitwidth analysis approach in PolyMage
compiler infrastructure \cite{mullapudi2015asplos, chugh16pact}. With the PolyMage DSL, FPGAs are targeted by
first generating High-Level Synthesis (HLS) code after a realization of
several transformations for parallelization and reuse; the HLS code is
subsequently processed by a vendor HLS suite (Xilinx Vivado in the case of
PolyMage). There have been several recent DSL efforts that target FPGAs for image
processing; these include Darkroom~\cite{darkroom},
Rigel~\cite{rigel}, Halide~\cite{halide17hls}, HIPAcc~\cite{hipacc16fpga} and
PolyMage-HLS~\cite{chugh16pact}. While these works
have addressed several challenges in compiling DSL to FPGAs, none of them
have studied the issue of exploiting application-dependent
variable fixed-point data types for power and area savings. HiPAcc
goes to the extent of providing pragmas for specifying bitwidths of
variables, but no automatic compiler support for it.

\begin{table}[t]
\begin{minipage}[b]{0.38\linewidth}
\includegraphics[scale = 0.8]{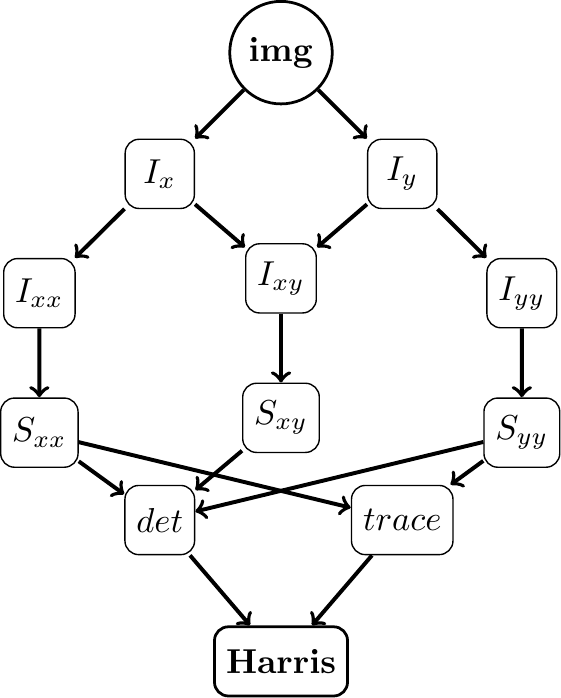}
\captionof{figure}{DAG representation of the Harris Corner Detection (HCD) algorithm.}
\label{fig:hdag}
\end{minipage}\hfill
\begin{minipage}[b]{0.58\linewidth}
\renewcommand{\arraystretch}{1.1}
\begin{tabular}{r  |r} 
\toprule
Stage  & Computation \\  
\hline 
$I_x$ &  $\frac{1}{12}\left[ \begin{smallmatrix} -1 & 0 & 1 \\ -2 & 0 & 2 \\ -1 & 0& 1\end{smallmatrix} \right]$  \\ 
$I_y$ &  $\frac{1}{12}\left[ \begin{smallmatrix} -1 & -2 & -1 \\ 0 & 0 & 0 \\ 1 & 2& 1\end{smallmatrix} \right]$  \\
$I_{xx}$ &  $I_x\left(i,j\right) I_x\left(i,j\right)$ \\
$I_{xy}$ &  $I_x\left(i,j\right) I_y\left(i,j\right)$  \\
$I_{yy}$ &  $I_y\left(i,j\right) I_y\left(i,j\right)$  \\
$S_{xx}$ &  $A=\left[ \begin{smallmatrix} 1 & 1 & 1 \\ 1 & 1 & 1 \\ 1 & 1 & 1\end{smallmatrix} \right]$\\
$S_{xy}$ &  $A$\\
$S_{yy}$ &  $A$\\
$det$ &  $S_{xx}\left(i,j\right) S_{yy}\left(i,j\right)-S_{xy}\left(i,j\right) S_{xy}\left(i,j\right)$\\
$trace$ &  $S_{xx}\left(i,j\right)+S_{yy}\left(i,j\right)$\\
$Harris$ &  $det\left(i,j\right) - 0.04 ~trace\left(i,j\right) trace\left(i,j\right)$\\ [1ex] 
\hline 
\end{tabular}
\caption{Summary of computations in HCD benchmark.}
\label{tab:hcd}
\end{minipage}
\end{table}

The rest of this paper is organized as follows. Related work is discussed in
Section~\ref{sec:related-work} and the necessary background is provided in
Section~\ref{sec:background}. Section~\ref{sec:bitwidth} presents in detail the
main contributions of this paper. 
Experimental evaluation is
presented in Section~\ref{sec:experiments} and  conclusions are presented in
Section~\ref{sec:conclusions}.

\section{Related Work}
\label{sec:related-work}
Besides PolyMage~\cite{mullapudi2015asplos},  Rigel~\cite{rigel},
Darkroom~\cite{darkroom}, HIPAcc~\cite{membarth16tpds}, and
Halide~\cite{kelley13pldi} are other recent domain-specific languages (DSL) 
for image processing pipelines.  Among them, PolyMage, Rigel, HIPAcc, and
Darkroom compilers can generate hardware designs targeting FPGAs, and none of
these currently optimize designs using bitwidth analysis.

There are several works on bitwidth estimation in digital signal processing
applications ~\cite{TEMA352, vakili13tcad, cong09fccm, lee06tcad,
zhang10jsip}. However, these techniques are not scalable and can only be applied to small circuits
like low degree polynomial multiplications, 8x8 discrete cosine transform
computation etc. which contain very few signals in the order of 10s and 100s.
Whereas the techniques proposed in this paper exploits both the image processing
domain and the PolyMage-HLS compilation framework to do interval analysis on
large image processing pipelines wherein each pixel at every stage of the pipeline
constitutes a signal. Further, there are a class of iterative algorithms such as
optical flow wherein errors in range estimation accumulate across iterations making
the analysis in essence useless. In this work, we show
how we can use SMT solvers to get accurate range estimates and thus contain errors
across iterations. Kinsman and Nicolici~\cite{kinsman10} proposed a SMT solver based
approach, however, they evaluated their approach on small signal processing
applications involving less than 10 signals. The SMT solver based approach proposed 
in the current work handles large image processing applications which are iterative
in nature with potentially thousands of pixel signals being processed in each 
iteration.

Usually, range analysis (integer bits) and precision (fraction bits)
analysis are performed separately. For precision analysis, there are
heuristic search~\cite{vakili13tcad, nguyen11precision} based approaches which try to minimize circuit area and power while satisfying  constraints on Signal-to-Noise 
ratio. The time complexity of these algorithms is usually very high and hence 
impractical to use in large image processing pipelines. Whereas, the greedy heuristic 
algorithm we proposed in this paper runs in linear time with respect to the number of stages
present in the image processing pipeline and is independent of the image dimensions.
Overall, ours is the first extensive study on the
application of practical range and precision analyses in image processing 
applications, and their impact on power and area savings.

Mahlke et al.~\cite{mahlke01tcad} proposed a data flow analysis based approach
for bitwidth estimation of integral variables in the PICO (Program-in Chip-out)
system for synthesizing hardware from loop nests specified in C.  Along the same
lines, Gort and Anderson~\cite{anderson13range} proposed a range analysis
algorithm in the LegUp HLS tool. Their  range analysis algorithm is designed
over the LLVM intermediate representation and is implemented as an LLVM 
analysis pass. On the other hand, the interval arithmetic based range 
analysis algorithm we will propose works at
the DSL level and furthermore, the  proposed compilation framework permits 
the
usage of any other range analysis algorithm nearly in a plug-and-play 
manner; this can otherwise require significant effort in order to make it 
into a compiler analysis pass.

The integral bitwidth analysis algorithm due to Budiu et 
 al.~\cite{budiu00europar} is similar to the previous work but uses a
 different data flow analysis formulation.
Stephenson et al.~\cite{stephenson00sigplan} performs integer bitwidth
 analysis through range propagation, again using a data flow analysis
 framework.
 Tong et al. \cite{tong00customfp} proposed the usage of variable
 bitwidth floating-point units, which can save power for applications that
 do not require the full range and precision provided by the standard
 floating-point data type.
  Sampson et al. \cite{sampson11enerj} proposed
 EnerJ, an extension to Java that supports approximate data types and
 computation.  However, fixed-point data types and the associated approximate
 operations are not considered in EnerJ. On the contrary, PolyMage DSL can be
 enhanced by using the approximate data types as proposed by EnerJ.

Approximate computing has a rich body of
literature~\cite{mittal16approxsurvey, ramani15approx, han13approx}.
However, our context of
domain-specific automatic HLS compilation is unique. Depending on the output
quality and the application in question, our approach could either be seen
as exploiting customized precision or leveraging approximate computing.
In addition to customized precision, we can potentially use approximate
arithmetic operations \cite{kahng12adder, liu14multiplier} in the various
stages of computation.

\section{Background}
\label{sec:background}
\begin{figure}[tb]
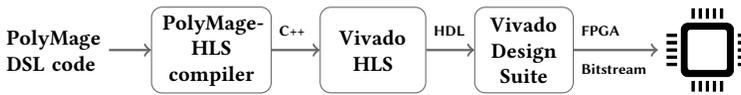

    \centering
        \overview
\vskip 5pt
\caption{PolyMage high-level compilation for FPGAs.}
\label{fig:overview}
\end{figure}
In this section, we briefly explain the architecture of the PolyMage-HLS
compilation framework; and further introduce the basics of interval and
affine arithmetic necessary to understand how range analysis techniques based
on them can be seamlessly integrated into our PolyMage-HLS compiler.
\subsection{PolyMage DSL and Compilation Framework}
In this paper, we use the PolyMage DSL and its compiler infrastructure
to implement and evaluate our automatic bitwidth analysis approach.
The PolyMage compiler infrastructure, when it was first
proposed~\cite{mullapudi2015asplos}, comprised an optimizing
source-to-source translator that generated OpenMP C++ code from an input
PolyMage DSL program. Chugh et al.~\cite{chugh16pact} developed a
backend for PolyMage targeting FPGAs by generating
HLS C++ code. The generated C++ code is translated into a hardware design 
expressed in a Hardware Description Language (HDL) such as VHDL or Verilog
using High Level Synthesis compiler.
Figure~\ref{fig:overview}
shows the entire design flow using the PolyMage-HLS compiler and Xilinx Vivado
tool chain. For syntactic details and code examples, we refer the reader to
the PolyMage webpage~\cite{polymage-web}.

\subsection{Interval and Affine arithmetic}
With interval analysis, one estimates the range of an output
signal $z \leftarrow f\left(x, y\right)$ based on the range of the input signals $x$
and $y$, and the function $f$. For example, if the range of $x$ and $y$ are
$[\low{x},\high{x}]$ and $[\low{y}, \high{y}]$ respectively, and
$z\leftarrow x+y$, then the range of $z$ is
$[\low{x}+\low{y}, \high{x} +\high{y}]$. Such range estimation functions have
to be defined for different operations that are applied iteratively to
obtain the ranges of different intermediate and output signals involved in
the computation. Although interval arithmetic is simple and easy
to use in to practice, it suffers from the problem of range over-estimation.
For example, if the range of a signal $x$ is $[5, 10]$, then the interval
arithmetic estimates the range of the expression $x - x$ as $[-5, 5 ]$ whereas
the actual range is $[0, 0]$. This is due to the fact that the interval
arithmetic ignores the correlations between the operand signals if there were any.

With affine arithmetic analysis, a signal $x$ is
represented in an affine form as $x=x_0 + \sum_{i=1}^{n} x_i \epsilon_i$
where $\epsilon_i \in [-1, 1]$ are interpreted as independent noise signals
and their respective coefficients $x_i$'s are treated as the weights
associated with them. The interval of the signal $x$ from its affine form
can be inferred as $[x_0-r, x_0+r]$ where $r=\sum_{i=1}^{n} |x_i|$. The
 addition
and subtraction operations on two input signals is defined as
$z=x\pm y = \left(x_0 \pm y_0\right) + \sum_{i=1}^{n}
\left(x_i \pm y_i\right)\epsilon_i$ and yields the resulting signal in its affine form.
The correlation between the signals
$x$ and $y$ is captured by sharing the independent noise signals $\epsilon_i,
1\leq i \leq n $ in their affine forms either partially or totally.
Now, when we perform a computation $x-x$ by considering the signal $x$ in its
affine form, the resulting range will be zero as against the over-estimated
range which we get in interval arithmetic analysis. Thus the techniques based
on affine arithmetic
arrive at better bounds when compared with interval analysis based techniques by
taking into account cancellation effects in computations involving
correlated signals.
However, note that if the operation is multiplication,
then the resulting signal contains quadratic terms and hence has to be
approximated to an affine form. A detailed discussion on affine arithmetic
analysis is beyond the scope of this paper and we recommend the reader to
Stolfi and Figueiredo~\cite{TEMA352} for the same.

\section{Bitwidth Analysis}\label{sec:bitwidth}
In this section, we present the main technical contributions of this paper which
are summarized below.
\begin{enumerate}
\item A simple interval arithmetic based range analysis algorithm
illustrating how DSLs facilitate practical and efficient program analysis
techniques when compared with C/C++ kind of languages
(Section~\ref{sec:ra}).
\item  A software architecture for range analysis in DSL compilers in which variants of interval and affine arithmetic based approaches
can be easily deployed (Section~\ref{sec:ca}).
\item An SMT solver based approach for range analysis
which substantially improves the accuracy of range estimates and contains the
propagation of estimation errors across iterations. Again, such an SMT solver 
based approach would have been hard to realize if not for the DSL compiler 
framework (Section~\ref{sec:smt}).
\item A profile driven approach for range analysis (Section~\ref{sec:profile}).
\item A greedy heuristic search technique for precision estimation
(Section~\ref{sec:precision}).
\end{enumerate}

\subsection{Variable Width Fixed-Point Data Types}
A fixed-point data type is specified by a tuple $(\alpha, \beta)$ where
$\alpha$ and $\beta$ denote the number of bits  allocated for representing
the integral and fractional parts respectively. The total bitwidth of the
data type is $\alpha + \beta$. The decimal value associated with a
fixed-point binary number $x=b_{\alpha-1}\ldots b_0 . b_{-1} \ldots
b_{-\beta}$ depends on whether it is interpreted as an unsigned integer or a
two's complement signed  integer, and is given as follows:
\begin{equation*}
value(x) = \begin{cases}
\sum^{\alpha-1}_{i=-\beta} 2^i b_i & \text{unsigned} \\
-2^{\alpha-1} b_{\alpha-1} + \sum_{i=-\beta}^{\alpha-2} 2^i b_i & \text{2's
    complement.}
\end{cases}
\end{equation*}
This gives us the ranges $[0, 2^{\alpha}-2^{-\beta}]$ and $[-2^{\alpha-1} ,
2^{\alpha-1}-2^{-\beta}]$ for unsigned and signed fixed-point numbers
respectively. The parameter $\alpha$ gives the range of values that can be
represented and the parameter $\beta$ indicates that the values in the range
can be represented at a resolution of $2^{-\beta}$. Hence, the range and
precision can be improved, by increasing $\alpha$ and $\beta$ respectively.
In this paper, we overload the term precision to also mean the entire data
type $(\alpha, \beta)$, and this can be disambiguated based on context.

Fixed-point data types are useful in image processing applications where the
range of values produced during computations is usually limited
and the precision requirements are less demanding  when compared to many
other numerical algorithms. The data type (range and precision) requirement
at a stage depends on the input data type and the nature of local
computations carried out at that particular stage. Further, overflows during
computations can be addressed by using saturation mode arithmetic instead of
the conventional wrap around arithmetic operations performed in CPUs and GPUs.
 The complexity of arithmetic operations on
fixed-point data type $(\alpha, \beta)$ is very similar to that of integer
operations on bitwidth $\alpha+\beta$.

\subsection{Range ($\alpha$) Analysis Algorithm}\label{sec:ra}
The number of integer bits required at a stage $I$ denoted as $\alpha_I$ is
a direct function of the bitwidth of the input data and the operations it
performs on them. The input data here refers to the data supplied to the
stage by its predecessor stages in the DAG. Further, the computations on
the pixel signals at each stage of DAG are identical and hence their
corresponding ranges would be the same. This information is implicitly provided
by a PolyMage DSL program and is hard to elicit from C like programs.
We use this insight to group all the pixel signals at a stage and perform a
combined range analysis using interval arithmetic. If the range of the
data produced at a stage is $[\low{x}, \high{x}]$, then the number of
integral bits $\alpha_I$ required to store the data without overflow is as
follows:
\begin{equation*}
    \alpha=
\begin{cases}
    \max(\ceil{\log_2(\ceil{|\low{x}|})},\ceil{\log_2(\floor{|\high{x}|}+1)})+1 & \text{if } \low{x} <0\\
    \ceil{\log_2(\floor{\high{x}}+1)},              & \text{otherwise}.
\end{cases}%
\end{equation*}
The number of fractional bits
$\beta_I$ required at a stage depends on the application-specific error
tolerance or quality metric, and we propose a profile-driven estimation
technique in Section~\ref{sec:profile}.

The range analysis algorithm iterates over the stages of a DAG in a
topologically sorted order. At each stage, an equivalent expression tree
for the computations (point-wise or stencil) is built. Then the range of the
pixel signals is estimated by recursively performing interval arithmetic on the
expression tree using one of following five interval arithmetic rules.
\begin{enumerate}
\item {$\mathbf{z=x+y:}$}  $[\low{z},\high{z}] = [\low{x}+\low{y}, \high{x}+\high{y}]$
\item {$\mathbf{z=x-y:}$}  $[\low{z},\high{z}] = [\low{x}-\high{y}, \high{x}-\low{y}]$
\item {$\mathbf{z=x*y:}$} $[\low{z},\high{z}] = [t_1, t_2 ]$  where \\$t_1 = \min(\low{x}\low{y}, \low{x}\high{y}, \high{x}\low{y}, \high{x}\high{y})$
and \\ $t_2 = \max(\low{x}\low{y}, \low{x}\high{y}, \high{x}\low{y}, \high{x}\high{y})$.
\item $\mathbf{z=x/y:}$$[\low{z},\high{z}] = [\low{x}, \high{x}] * [1/\high{y}, 1/\low{y}]$ if $0 \not \in [\low{y}, \high{y}]$
\item $\mathbf{z = x^n}$
\begin{enumerate}
  \item $n$ is odd: $[\low{z},\high{z}] = [\low{x}^n, \high{x}^n ]$
  \item $n$ is even :
  $\begin{aligned}[t]
  [\low{z},\high{z}] & = & [\low{x}^n, \high{x}^n ] ~\textrm{if}~\low{x} \geq 0 \\
                      & = & [\high{x}^n, \low{x}^n ] ~\textrm{if}~\high{x} < 0\\
                      & = & [0, max\{\low{x}^n, \high{x}^n\} ]~\textrm{otherwise}
  \end{aligned}$
\end{enumerate}
\end{enumerate}

\subsection{Bitwidth Analysis Compilation Framework}\label{sec:ca}
The interval arithmetic based range analysis algorithm proposed in the
previous section uses the fact that all the pixel signals in each stage of an
image processing DAG are homogeneous in nature and groups them to do a
combined range analysis. However, other analysis techniques such as those
based on affine arithmetic cannot be applied on the DSL level programs in the
same fashion. In this section, we show how interval and affine arithmetic based 
range analysis techniques can be deployed with ease in the PolyMage compilation 
framework.

Recall that the PolyMage-HLS compiler translates a DSL program into C++ code
which the Xilinx Vivado HLS compiler synthesizes into an equivalent circuit
for a target FPGA.  For example, Listing~\ref{lst:sobel} depicts the C++ code
generated by the PolyMage-HLS compiler when Sobel-x filter is applied on an
input image. The generated C++ program can be run in a purely simulation
mode after compilation on any processor by providing test input images as
stimulus. It can be noted from the Listing~\ref{lst:sobel}, that the data type
of the stream, line and window buffers are parameterized by the type
{\tt\bf typ}. It can be a float or any fixed-point data type $(\alpha, \beta)$.
During the hardware synthesis or in the simulation mode, using the C++
polymorphism feature, corresponding libraries for the arithmetic operations
will be invoked based on the operand types.
Now, the parameter {\tt\bf typ} can also be set to an interval type which is
defined in a suitably chosen interval analysis library. If the generated C++
program contains a statement $x=y+z$, then depending on the type of the
variables $x$, $y$ and $z$ (like float, ia-type, aa-type etc.), appropriate
addition operation will be invoked. For example, in order
to perform affine arithmetic analysis on the Sobel-x program, using the
Yet Another Library for Affine Arithmetic (YalAA)~\cite{yalaa}, all we have to
do is to define the parameter {\tt\bf typ} appropriately as depicted in
Listing~\ref{lst:yalaa}.
\begin{table}[t]
    \noindent\begin{minipage}[t]{.47\textwidth}
    \begin{lstlisting}[language=C,label={lst:sobel},
    breaklines=true,
    basicstyle=\footnotesize,
    numbers=left, 
    keywordstyle=\sffamily\bfseries\color{green!40!black}, commentstyle=\itshape\color{purple!40!black},
    morekeywords={typ}, identifierstyle=\color{blue},frame=single,%captionpos=b,
        caption=Auto-generated restructured HLS code for Sobel-x.]
    #include <hls_stream.h>
    #include <malloc.h>
    #include <cmath>
    #include <arith.h>
    
    void sobel_x(hls::stream<typ> & img, 
                 hls::stream<typ> & sobel_x_out) 
    {
      const int  _ct0 = (2 + R);
      const int  _ct1 = (2 + C);
      hls::stream<typ> Ix_out_stream;
      hls::stream<typ> img_Iy_stream;
      typ  Ix_img_LBuffer[3][_ct1];
      typ  Ix_img_WBuffer[3][3];
      typ  Ix_img_Coeff[3][3];
    
      /* Code for Sobel-x stage follows. */
    }
    \end{lstlisting}
    \end{minipage}\hfill
      \begin{minipage}[t]{.47\textwidth}
    \begin{lstlisting}[language=C,label={lst:yalaa},breaklines=true,
    basicstyle=\footnotesize\ttfamily,keywordstyle=\sffamily\bfseries\color{green!40!black}, commentstyle=\itshape\color{purple!40!black},
    numbers=left, 
    morekeywords={typ}, identifierstyle=\color{blue}, frame=single,framexbottommargin=100pt,%captionpos=b,
        caption=Type definitions for Affine and Interval Analysis.]
    // Switch for Affine analysis 
    #ifdef AFFINE
    #include <yalaa.hpp>
    typedef yalaa::aff_e_d typ;
    #endif
    // Switch for Interval analysis 
    #ifdef INTERVAL
    #include <Easyval.hpp>
    typedef Easyval typ;
    #endif
    \end{lstlisting}
    \end{minipage}
    \end{table}
When we run
the generated C++ program with this data type definition, the value
associated with each pixel in each stage of the pipeline DAG is its affine
signal value which contains the base signal and the coefficients for the noise
variables. From this the range of every pixel can be derived as explained
before. If the data type corresponds to interval arithmetic, then the value
associated with each pixel is an interval. Using this approach, any kind
of interval analysis technique can be deployed in the PolyMage-HLS compiler
easily without re-architecting the analysis backend.

In the next section, we show how interval arithmetic based techniques can
fare really poorly if the benchmarks contain certain kinds of computational
patterns; we use an Optical Flow algorithm as an example. We then  propose our new range analysis technique using SMT solvers.

\subsection{Range Analysis using SMT Solvers} \label{sec:smt}
\begin{table*}[t]
\begin{minipage}[b]{0.38\linewidth}
\centering
\includegraphics[scale = 0.6]{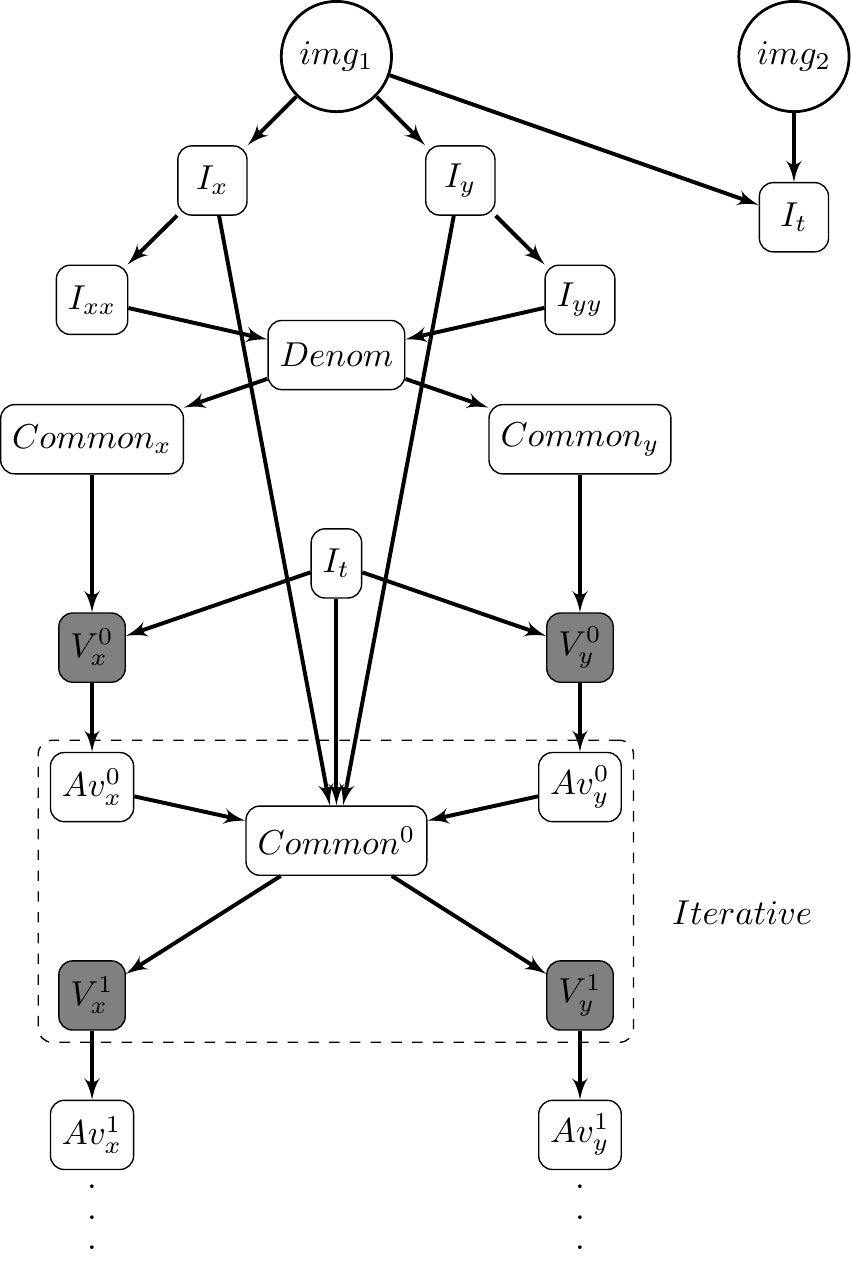}
\captionof{figure}{DAG representation of the Horn-Schunck Optical Flow algorithm.}
\label{fig:hsof_dag}
\end{minipage}
\hfill
\begin{minipage}[b]{0.58\linewidth}
\centering 
\resizebox{\linewidth}{!}{%
\renewcommand{\arraystretch}{1.1}
\begin{tabular}{r |r} 
\toprule
Stage  & Computation \\  
\midrule
$I_{t}$     &  $img_1\left(i,j\right)-img_2\left(i,j\right)$ \\
$I_x$ 	    &  $\frac{1}{12}\left[ \begin{smallmatrix} -1 & 0 & 1 \\ -2 & 0 & 2 \\ -1 & 0& 1\end{smallmatrix} \right]$  \\ 
$I_y$ 	    &  $\frac{1}{12}\left[ \begin{smallmatrix} -1 & -2 & -1 \\ 0 & 0 & 0 \\ 1 & 2& 1\end{smallmatrix} \right]$  \\
$I_{xx}$    &  $I_x\left(i,j\right) I_x\left(i,j\right)$ \\
$I_{yy}$    &  $I_y\left(i,j\right) I_y\left(i,j\right)$ \\
$denom$     &  $\alpha^2+I_{xx}\left(i,j\right)+I_{yy}\left(i,j\right)$ \\
$Common_x$  &  $\frac{I_x\left(i,j\right)}{Denom\left(i,j\right)}$ \\
$Common_y$  &  $\frac{I_y\left(i,j\right)}{Denom\left(i,j\right)}$ \\
$V_x^0$      &  $-I_t\left(i,j\right)Common_x\left(i,j\right)$ \\
$V_y^0$      &  $-I_t\left(i,j\right)Common_y\left(i,j\right)$ \\
$Av_x^0$     &  $A=\frac{1}{4} \left[ \begin{smallmatrix} 0 & 1 & 0 \\ 1 & 0 & 1 \\ 0 & 1 & 0\end{smallmatrix} \right]$\\
$Av_y^0$     &  $A$\\
$Common^0$  &  $I_x(i,j)Av_x^0(i,j)+  I_y(i,j)Av_y^0(i,j)  + I_t(i,j) $ \\
$V_x^1$      &  $Av_x^0(i,j)- Common^0(i,j)Common_x(i,j))$ \\
$V_y^1$      &  $Av_y^0(i,j)-Common^0(i,j)Common_y(i,j))$ \\
\midrule
\end{tabular}
}
\caption{Summary of computations in the Optical Flow algorithm.}
\label{tab:of_comp}
\end{minipage}
\end{table*}

Range analysis algorithms based on interval or affine arithmetic variants 
have
limitations in capturing the correlations between computations 
(refer Section~\ref{sec:background}). For example, consider the
Optical Flow benchmark, whose DAG and computations at
each stage of the DAG are given in the Figure~\ref{fig:hsof_dag} and 
Table~\ref{tab:of_comp} respectively. Consider the point-wise stage 
$Common_x$ where each pixel is computed as follows:
\[
Common_x(i,j) = \frac{I_x(i,j)}{Denom(i,j)}.
\]
The second column in Table~\ref{tab:ofrange} represents the ranges inferred at various
stages of the Optical Flow benchmark using interval analysis. The ranges obtained
using affine analysis are also very similar with no change in bitwidth estimates.
We observe that the range
at $Common_x$ stage is inferred as $[-21.25, 21.25]$ and hence requires
6 integral fixed-point bits. This range is obtained by dividing the range 
of $I_x$ with the range of $Denom$, which are $[-85, 85]$ and $[4, 14454]$ respectively.  However, if we symbolically expand the computation at the
stage $Common_x$, then we obtain the following formula:
\begin{eqnarray}
Common_x(i,j) = \frac{I_x(i,j)}{\alpha^2+I_x(i,j)I_x(i,j)+I_y(i,j)I_y(i,j)}.
\label{eq:eq1}
\end{eqnarray}
Now, we observe that the pixel signal $I_x(i,j)$ is present both in the
numerator and denominator. Since $\alpha=2$, the RHS in 
Equation~\eqref{eq:eq1} is equivalent to $\frac{x}{x^2+a}$ for some $a\geq 4$.
Figure~\ref{fig:commonplt} shows the plot of the 
function $\frac{x}{x^2+a}$ for various values of $a$.
We can analytically determine the absolute values of the maximum and minimum
of that function to be less than one.  Both interval and affine
arithmetic analysis fail to arrive at this conclusion. 
\begin{table}[htb]
\footnotesize
    \centering
    \caption{Comparison of range estimates using interval analysis and SMT 
    solver based approach for the Optical Flow benchmark. Integral bitwidths 
    ($\alpha^{IA}$ and $\alpha^{\smtra}$) derived from the range estimates 
    are also provided. The fractional bitwidth estimates ($\beta$) from profile-guided
    heuristic search are also provided for completeness.  
    \label{tab:ofrange}}
\resizebox{\textwidth}{!}{
\begin{tabularx}{1.3\textwidth}{lcc|cc|c|c}
    \toprule
    \multirow{2}{*}{Stage} & \multicolumn{2}{c|}{Interval Analysis}      & \multicolumn{2}{c|}{Z3RA Analysis} & \multirow{2}{*}{$\beta$-Bitwidth} & \\
                                 & $\alpha^{IA}$-Range & $\alpha^{IA}$-Bitwidth & Z3RA-Range & Z3RA-Bitwidth &                    & \\  
    \midrule
    Img$_1$, Img$_2$                   & $(0,255)$                                  & 8  & $(0,255)$                &  8     & 0        &\\
    I$_t$                              & $(-255,255)$                               & 9  & $(-254.87 , 254.99)$     &  9     & 0        &\\
    I$_x$,I$_y$                        & $(-85,85)$                                 & 8  & $(-84.91 , 84.88)$       &  8     & 9        &\\
    I$_{xx}$,I$_{yy}$                  & $(0,7225)$                                 & 14 & $(0 , 7210.01)$          & 13     & 3        &\\
    Denom                              & $(4,14454)$                                & 15 & $(4, 14423.97)$          & 14     & 3        &\\
    \pbox{0.02cm}{$Common_x$, \\ $Common_y$} & $(-21.25,21.25)$                     & 9        & $(-0.22 , 0.19)$   &  1     & 9        &\\
    V$_{x}^0$,V$_{y}^0$                & $(-5418.75,5418.75)$                       & 8  & $(-56.09 , 56.07)$       &  7     & 8        &\\
    \midrule
    Avg$_x^0$,Avg$_y^0$                & $(-5418.75,5418.75)$                       & 7  & $(-56.09 , 56.07)$       &  7     & 8        &\rdelim\}{3}{20pt}$(stage 1)$\\
    Common$^0$                         & $(-921443, 921443)$                        & 13 & $(-9778.67 , 9781.98)$   & 14     & 3        &\\
    V$_x^1$,V$_y^1$                    & $(-1.95861 * 10^7, 1.95861 * 10^7)$        & 13 & $(-103.10 , 102.41)$     &  9     & 7        &\\
    \midrule
    Avg$_x^1$,Avg$_y^1$                & $(-1.95861 * 10^7, 1.95861 * 10^7)$        & 10 & $(-158.54 , 158.00)$     &  9     & 9        &\rdelim\}{3}{20pt}$(stage 2)$\\
    Common$^1$                         & $(-3.32964 * 10^9, 3.32964 * 10^9)$        & 18 & $(-22464.30 , 22471.98)$ & 16     & 4        &\\
    V$_x^2$,V$_y^2$                    & $(-7.07743 * 10^{10}, 7.07743 * 10^{10})$  & 18 & $(-295.04 , 294.35)$     & 10     & 8        &\\
    \midrule
    Avg$_x^2$,Avg$_y^2$                & $(-7.07743 * 10^{10}, 7.07743 * 10^{10})$  & 18 & $(-295.04 , 294.35)$     & 10     & 9        &\rdelim\}{3}{20pt}$(stage 3)$\\
    Common$^2$                         & $(-1.20317 * 10^{13}, 1.20317 * 10^{13})$  & 25 & $(-39349.61 , 39363.12)$ & 17     & 5        &\\
    V$_x^3$,V$_y^3$                    & $(-2.55743 * 10^{14}, 2.55743 * 10^{14})$  & 25 & $(-492.36 , 491.17)$     & 10     & 9        &\\
    \midrule
    Avg$_x^3$,Avg$_y^3$                & $(-2.55743 * 10^{14}, 2.55743 * 10^{14})$  & 25 & $(-492.36 , 491.17)$     & 10     & 9        &\rdelim\}{3}{20pt}$(stage 4)$\\
    Common$^3$                         & $(-4.34763 * 10^{16}, 4.34763 * 10^{16})$  & 33 & $(-64412.09 , 64434.17)$ & 17     & 7        &\\
    V$_x^4$,V$_y^4$                    & $(-9.24127 * 10^{17}, 9.24127 * 10^{17})$  & 33 & $(-794.02 , 792.90)$     & 11     & 9        &\\
    \bottomrule
\end{tabularx}
}
\end{table}

We address this issue using an SMT solver based range analysis approach.  
The basic idea is to build a constraint system involving the variables 
$I_x(i,j)$,
$I_y(i,j)$ and $Common_x(i,j)$. The constraint system consists of range 
constraints on variables $I_x(x,y)$ and $I_y(i,j)$, that are inferred 
through
interval analysis, and an equality constraint as specified in the 
Equation~\eqref{eq:eq1}. To this base constraint system, we add a parametric
constraint $Common_x(i,j) >UB$, $UB$ being the parameter. For a given value 
of $UB$, if the constraint system has no solution, then we know that the 
maximum
value of $Common_x(i,j)$ is bounded by UB. We use this idea to arrive at a
tight upper bound using a binary search algorithm. The upper bound estimate
need not be too accurate as long as it does not affect the corresponding bit 
width estimates. A similar approach is adopted to determine the lower
bound too.

We observe from the following recurrence relations that a bad estimate
in the bitwidth of stages $Common_x$ and $Common_y$ has a cascading effect 
on the bitwidth estimates
of stages $V_x^{k}$, $V_y^{k}$, $Av_x^k$, $Av_y^{k}$ and $Common^k$ for 
any $k\geq 0$:
\begin{eqnarray*}
V_x^0(i,j) & = & -I_t(i,j)Common_x(i,j) \\
V_y^0(i,j) & = & -I_t(i,j)Common_y(i,j) \\
Av_x^k(i,j) & = & \frac{1}{4}\begin{bmatrix} 0 & 1 & 0 \\ 1 & 0 & 1 \\ 0 & 1 & 0 \end{bmatrix} \circledast \begin{bmatrix} V_x^k(i-1,j-1) & V_x^k(i-1,j) & V_x^k(i-1,j+1) \\ V_x^k(i,j-1) & V_x^k(i,j) & V_x^k(i,j+1) \\ V_x^k(i+1,j-1) & V_x^k(i+1,j) & V_x^k(i+1,j+1) \end{bmatrix}\\
Av_y^k(i,j) & = & \frac{1}{4}\begin{bmatrix} 0 & 1 & 0 \\ 1 & 0 & 1 \\ 0 & 1 & 0 \end{bmatrix} \circledast \begin{bmatrix} V_y^k(i-1,j-1) & V_y^k(i-1,j) & V_y^k(i-1,j+1) \\ V_y^k(i,j-1) & V_y^k(i,j) & V_y^k(i,j+1) \\ V_y^k(i+1,j-1) & V_y^k(i+1,j) & V_y^k(i+1,j+1) \end{bmatrix}\\
Common^k(i,j) & = & I_x(i,j)Av_x^k(i,j) + I_y(i,j)Av_y^k(i,j) + I_t(i,j) \\
V_x^{k+1}(i,j) & = & Av_x^k(i,j) - Common^k(i,j)Common_x(i,j) \\
V_y^{k+1}(i,j) & = & Av_y^k(i,j) - Common^k(i,j)Common_y(i,j). \\
\end{eqnarray*}
The stages $Av_x^k$ and $Av_y^k$ are stencil stages which average the
values from the stages $V_x^{k}$ and $V_y^{k}$ respectively. Hence, any
bitwidth overestimates at the stages $V_x^{k}$ and $V_y^{k}$ will be 
directly
passed down to the stages $Av_x^k$ and $Av_y^k$. These in turn will be
reflected in the bitwidth estimate of the stage $Common^k(i,j)$.  Finally,
while estimating the bitwidth at the stage $V_x^{k+1}(i,j)$ , the bitwidth 
estimate errors of the stages $Common^k(i,j)$ and $Common_x(i,j)$ add-up 
linearly. Similar is the case for the stage $V_y^{k+1}(i,j)$.
We observe from Table~\ref{tab:ofrange} and Figure~\ref{fig:bitsgraph} as to 
how bitwidth estimates explode with each  stage using interval analysis, 
while they are contained using SMT solver based
approach. In the next section, we provide a more detailed description of our 
SMT-based range analysis algorithm called \smtra.
\begin{figure}[t]
    \begin{minipage}[t]{0.49\linewidth}
    \centering
    \resizebox{\linewidth}{!}{
                    \begin{tikzpicture}[scale=0.6]
\begin{axis}[
	cycle list name=exotic,
	ylabel={\textbf{Bits required}},
	xlabel={\textbf{Iterative Stage}},
  xlabel style={inner sep=10mm},
  ymin=0, ymax=65,
  xmin=9, xmax=21,
  axis y line*=left,
  axis x line*=bottom,
  y axis line style={opacity=0.1},
  x axis line style={opacity=0.1},
  ymajorgrids=true,
  xmajorgrids=true,
  legend style=
  {
  	draw=none,
  	at={(0.2,1)},
   	anchor=north,
   	legend columns=1,
  	/tikz/every even column/.append style={column sep=0.05cm}
 	},
  y tick label style={font=\small},
 	x tick label style={font=\small, rotate=50, anchor=north east, inner sep=0mm},
  xtick={9,10,11,12,13,14,15,16,17,18,19,20,21},
  xticklabels={$V_x^0$, $Av_x^0$, $Common^0$, $V_x^1$, $Av_x^1$, $Common^1$, $V_x^2$, $Av_x^2$, $Common^2$, $V_x^3$, $Av_x^3$, $Common^3$, $V_x^4$},
  ytick={0,10,20,30,40,50,60},
]

\addplot+[line width=1.1pt] plot coordinates {
  (9 , 14)
  (10, 14)
  (11, 21)
  (12, 26)
  (13, 26)
  (14, 33)
  (15, 38)
  (16, 38)
  (17, 45)
  (18, 49)
  (19, 49)
  (20, 57)
  (21, 61)
};
\addlegendentry{$\alpha^{IA}$}

\addplot+[line width=1.1pt] plot coordinates {
  (9 , 7)
  (10, 7)
  (11, 14)
  (12, 9)
  (13, 9)
  (14, 16)
  (15, 10)
  (16, 10)
  (17, 17)
  (18, 10)
  (19, 10)
  (20, 17)
  (21, 11)
};
\addlegendentry{$\alpha^{Z3RA}$}

\draw [decoration={brace},decorate,opacity=0.5] 
  (axis cs:10,0.5) --
    node[above=1pt,font=\scriptsize] {Stage 1} 
  (axis cs:12,0.5);
\draw [decoration={brace},decorate,opacity=0.5] 
  (axis cs:13,0.5) --
    node[above=1pt,font=\scriptsize] {Stage 2} 
  (axis cs:15,0.5);
\draw [decoration={brace},decorate,opacity=0.5] 
  (axis cs:16,0.5) --
    node[above=1pt,font=\scriptsize] {Stage 3} 
  (axis cs:18,0.5);
\draw [decoration={brace},decorate,opacity=0.5] 
  (axis cs:19,0.5) --
    node[above=1pt,font=\scriptsize] {Stage 4} 
  (axis cs:21,0.5);

\end{axis}
\end{tikzpicture}
                \label{fig:bitgraph}
                }
    \caption{Comparison of bitwidths obtained by SMT and Interval Analysis.}
    \label{fig:bitsgraph}
    \end{minipage}%
    \hfill
    \begin{minipage}[t]{0.49\linewidth}
    \resizebox{\linewidth}{!}{
            \centering
    \begin{tikzpicture}
	\begin{axis}[
                cycle list name=exotic,
    			axis lines = left,
                ymax=0.010,
    			xlabel={$I_x$},
    			ylabel={$Common_x$},
    			legend style={at={(0.05,0.9)},anchor=west},
    			xlabel style={inner sep=10mm},
    			x tick label style={font=\small},
    			y tick label style={font=\small}
	]
	\addplot+[
    domain=-85:85, 
    samples=10,
    smooth 
	]
	{x/(4 + x^2 + 3600)};
	\addlegendentry{$I_x/(4 + I{_x}^2 + 60^2)$}
	\addplot+[
    domain=-85:85, 
    samples=10, 
	]
	{x/(4 + x^2 + 7225)};
	\addlegendentry{$I_x/(4 + I{_x}^2 + 85^2)$}
	\end{axis}
	\end{tikzpicture}
    }
    \caption{Plot for $Common_x$ stage for $I_y=60$ and $I_y=85$}
    \label{fig:commonplt}
    \end{minipage}
    \end{figure}
\subsubsection{\smtra \ Algorithm}
The range of a pixel signal $S_{ij}$ at a stage $S$ of the input DAG depends 
on the pixel signals from the predecessor stages. Let $Dep(S_{ij})$ denote 
the
pixel signals from the input stages on which $S_{ij}$ is dependent, i.e.,  
\[
Dep(S_{ij}) = \{~I(k,l)~|~\textrm{$I$ is an input stage and $S_{ij}$ depends
on the $(k,l)^{th}$ pixel of $I$}. \}
\]

Then we can compute $Dep(S_{ij})$ by applying one of the following three 
cases recursively:
\begin{enumerate}
    \item $S$ is an input stage with no predecessors. Then $Dep(S_{ij})=\{S_{ij}\}$.
    \item $S$ is a point-wise stage. Then 
    \[
    Dep(S_{ij}) = \bigcup\limits_{P\in Predecessor(S)} Dep(P_{ij})
    \]
    where $Predecessor(S)$ is the set of immediate predecessors of stage $S$.
    \item $S$ is a stencil stage. A stencil stage has only one predecessor 
    stage. Let $P$ be the predecessor stage of $S$ and 
    \[
    \hat{P}=\{~(k,l)~|~\textrm{$S_{ij}$ depends on $P_{kl}$}\}.
    \]
    Then,
    \[
    Dep(S_{ij}) = \bigcup\limits_{(k,l) \in \hat{P}} Dep(P_{kl}).
    \]
\end{enumerate}

\begin{figure}[!htb]
    \centering
\begin{minipage}[t]{0.45\linewidth}
    \centering
\resizebox{\linewidth}{!}{
        \begin{tikzpicture}[scale=.9,every node/.style={minimum size=1cm},on grid]
		
	\begin{scope}[
    	yshift=-150,every node/.append style={
    	    yslant=0.5,xslant=-1},yslant=0.5,xslant=-1
    	             ]
		\fill[white,fill opacity=.9] (0,0) rectangle (4.5,4.5);
        \draw[black, dashed] (0,0) rectangle (4.5,4.5);
        \draw[step=5mm, black, dashed] (0,0) grid (4.5,4.5);
        \fill[green!60!black, fill opacity=0.9] (2,2) rectangle (2.5,2.5);
    \end{scope}
    
    \begin{scope}[
    	yshift=0,every node/.append style={
    	    yslant=0.5,xslant=-1},yslant=0.5,xslant=-1
    	             ]
        \fill[white,fill opacity=.9] (0,0) rectangle (4.5,4.5);
        \draw[black, dashed] (0,0) rectangle (4.5,4.5);
        \draw[step=5mm, black, dashed] (0,0) grid (4.5,4.5);
        \fill[green!60!black, fill opacity=0.9] (1.5,1.5) rectangle (3,3);
    \end{scope}

    \begin{scope}[
    	yshift=150,every node/.append style={
    	    yslant=0.5,xslant=-1},yslant=0.5,xslant=-1
    	             ]
        \fill[white,fill opacity=.9] (0,0) rectangle (4.5,4.5);
        \draw[black, dashed] (0,0) rectangle (4.5,4.5);
        \draw[step=5mm, black, dashed] (0,0) grid (4.5,4.5);
        \fill[green!60!black, fill opacity=0.9] (1,1) rectangle (3.5,3.5);
    \end{scope}
    
    \draw[-latex,thick, dashed] (1.5,2.25)  -- (0.5,-3);
    \draw[-latex,thick, dashed] (0,3)       -- (0,-2.9);
    \draw[-latex,thick, dashed] (-1.5,2.25) -- (-0.45,-3);
    \draw[-latex, dashed] (0,1.55)    -- (0,-3.35);
    
    \draw[-latex,thick, dashed] (2.45,7.45)  -- (1.5,2.25);
    \draw[-latex,thick, dashed] (0,8.75)       -- (0,3);
    \draw[-latex,thick, dashed] (-2.45,7.5) -- (-1.5,2.25);
    \draw[-latex, dashed] (0,6.3)    -- (0,1.55);

    \draw[-latex,thick] (5.9,-2.5) node[right]{$S_2$}
         to[out=180,in=90] (3.5,-2.5);

    \draw[-latex,thick](5.9,2.8)node[right]{$S_1$}
        to[out=180,in=90] (3.5,2.8);

    \draw[-latex,thick](5.9,8)node[right]{$I$}
        to[out=180,in=90] (3.5,8);

\end{tikzpicture}
        }
    \caption{An output pixel from the 3x3 stencil stage $S_2$ depends on a 5x5 window of pixels from the input image. This dependency is induced via a 3x3 window of pixels from stage $S_1$. }
    \label{fig:grid}
\end{minipage}%
\hfill
\begin{minipage}[t]{0.45\linewidth}
    \centering
\resizebox{\linewidth}{!}{
    \begin{tikzpicture}[scale=0.6]
\begin{axis}[
	cycle list name=exotic,
	ylabel={\textbf{\% Pixels}},
	xlabel={\textbf{Bits required}},
        	ymin=0, ymax=100,
        	xmin=2, xmax=8,
        	axis y line*=left,
      	axis x line*=bottom,
      	y axis line style={opacity=0.1},
      	x axis line style={opacity=0.1},
    	ymajorgrids=true,
    	xmajorgrids=true,
       	legend style=
       		{
       		draw=none,
        	  	at={(0.7,0.1)},
         	   	anchor=south,
         	 	legend columns=1,
        	 	/tikz/every even column/.append style={column sep=0.05cm}
 	 	},
 	every tick label/.append style={font=\small},
        	xtick={2,3,4,5,6,7,8,9,10,11,12,13,14},
        	ytick={0,10,20,30,40,50,60,70,80,90,100},
]

\addplot+[line width=1.1pt] plot coordinates {
  (2, 56.14206429)
  (3, 72.70465)
  (4, 84.73822857)
  (5, 92.81682857)
  (6, 97.81227857)
  (7, 99.91657857)
  (8 ,100)
};
\addlegendentry{Stage $I_x$}

\addplot+[line width=1.1pt] plot coordinates {
  (2,  44.94556429)
  (3, 56.14206429)
  (4 ,64.29742143)
  (5 ,72.70465)
  (6 ,79.04923571)
  (7 ,84.73822857)
  (8 ,89.15143571)
  (9 ,92.81682857)
  (10,  95.65925714)
  (11 , 97.81227857)
  (12  ,99.26000714)
  (13  ,99.91657857)
  (14  ,100)
};
\addlegendentry{Stage $I_{xy}$}

\end{axis}
\end{tikzpicture}
\label{fig:hs1}
}
    \caption{Cumulative distribution of pixels with respect to maximum 
    integral bitwidth length at stages $I_x$ and $I_{xy}$ of the HCD 
    benchmark.}
    \label{fig:hs}
\end{minipage}
\end{figure}
The algorithmic plan is to take pixel signal $S_{ij}$ and express its
computation using the signals from the set $Dep(S_{ij})$. The set of
equations which leads to its computation defines a constraint system.
We augment this constraint system by adding interval constraints on
input pixel signals from $Dep(S_{ij})$. In order to estimate the upper bound,
we add a constraint $S_{ij} > UB$ where $UB$ is a large enough constant,
and check if there is a solution. If there is no solution, then $UB$ 
is in fact an upper bound on $S_{ij}$. We continue to tighten the upper bound
using binary search until it reaches a stage where any further
improvement results in no bitwidth savings. 

Although the proposed algorithmic plan is theoretically sound,  in practice, 
there will be an explosion in the number of variables in the constraint 
system due to the presence of stencil
stages in the computational paths. For example, if an input is supplied
to a stage $S$ through a pipeline path in the DAG that consists of $k$
stencil stages such as $Av_x$ (cf. Table~\ref{tab:of_comp}), then the number
of variables in the constraint system grows quadratically, i.e.,
$|Dep(S_{ij}|=\theta(k^2)$. Even the state-of-the-art
SMT solvers may not be able to solve such large constraint systems using 
reasonable computational power. Figure~\ref{fig:grid} illustrates this 
scenario. Here, $I$ is the input image, stages $S_1$ and $S_2$ are two 3x3 
stencil stages.  A pixel in stage $S_2$ depends on 9 pixel signals from 
$S_1$ which in turn leads to a dependence on 25 pixel signals from input 
$I$.

We circumvent this explosion of variables in the constraint system by 
limiting
the expansion of computation at a stencil stage. Towards this, we define a new
function $\widehat{Dep}(S_{ij})$ as follows.
\begin{enumerate}
    \item If $S$ is an input stage with no predecessors, 
    $\widehat{Dep}(S_{ij})=\{S_{ij}\}$.
    \item If $S$ is a point-wise stage, \[
    \widehat{Dep}(S_{ij}) = \bigcup\limits_{P\in Predecessor(S)} \widehat{Dep}(P_{ij})
    \]
    where $Predecessor(S)$ is the set of immediate predecessors of stage $S$.
    \item If $S$ is a stencil stage, \[
    \widehat{Dep}(S_{ij}) = \{S_{ij}\}. 
    \]
\end{enumerate}
For example, in the optical flow benchmark, while computing 
$\widehat{Dep}(Common_x)$, the recursion terminates with the stencil stages 
$I_x$ and $I_y$. The range at stencil stages is estimated using simple 
interval
analysis. In the constraint system associated with the range estimation 
of the stage $Common_x$,  we use the range constraints which are already derived
on the pixel signals from stages $I_x$ and $I_y$. Thus we contain the number 
of variables in the constraint system from growing exponentially. 

To summarize, we consider the nodes in the DAG in a topologically sorted 
order.  We estimate the range at a stencil stage using a simple interval
analysis. And at a point-wise stage $S$, we construct a pruned computational
DAG wherein the stage $S$ acts as a sink and the source nodes are either 
input stages or stencil stages from which there exists a stencil-free path 
to stage $S$. Then, the computation of a pixel signal from stage $S_{ij}$ is
expressed using pixel signals from the source and intermediate nodes. This
set of equations acts as a base constraint system to which we add the
range constraints on the source pixel signals from the set 
$\widehat{Dep}(S_{ij})$.  Then we search for a tight lower bound constraint, 
$LB \leq S_{ij}$, and an upper bound constraint, $S_{ij} \leq UB$, using 
binary search, leading to a range estimate $S_{ij}\in [LB, UB]$. 

In the next section, we present a profile-driven analysis that provides
a lower bound on the bitwidth estimates, and show in the experimental 
results section, that the bitwidth estimates derived from the SMT solver 
based approach match the lower bounds provided by profile-driven analysis.

\subsection{Profile-Driven Analysis}
\label{sec:profile}
Profile-driven analysis can be used to accomplish two tasks. First, we can obtain
lower bounds on the bitwidth estimates, which can be compared with the estimates
obtained using static analysis. Second, depending on the application, these 
estimates
can be used in the actual system design instead of the conservative 
estimates obtained through static analysis techniques. However, the
bitwidth  requirements estimated at each stage using profiling naturally 
depend on the sample input images. Based on the analysis done by Torralba et 
al. \cite{torralba2003statistics}, we hypothesize that the
images taken from a certain domain, like for example {\it nature}, has
similar properties, and hence the bitwidth estimates can be carried over to
other images drawn from the same domain.

\subsubsection{Integral Bits}
The number of integral bits required at a stage $i$ denoted as $\alpha_i$  can
be obtained by running the input PolyMage program on a sample  distribution of
input images. Let $\alpha_i^s$ be the maximum number of bits  required by stage
$i$ to represent a pixel from an image sample $s$. Then  the average number of
bits $\alpha_i^{avgP}$ required based on a sample set  $S$ is $\sum_{s\in S}
\alpha_i^s/|S|$. Similarly, the worst-case number of  bits $\alpha_i^{maxP}$
required is $max_{s\in S} \alpha_i^s$. We can either  use $\alpha_i^{avgP}$ or
$\alpha_i^{maxP}$ as estimates for $\alpha_i$. Even if the estimate does not
suit certain images, in many application  contexts, using saturation mode
arithmetic results in satisfying the desired  output quality metric. Let
$\alpha_i^{IA}$ and $\alpha^{\smtra}$ be the integral bitwidth  estimates
obtained for stage $i$ through interval analysis and \smtra\ analysis
respectively. For the benchmarks we have considered, affine analysis show some
improvements in the range estimates, but it amounts to same bitwidth requirement
as with interval analysis. Hence,  throughout the rest of the paper, we consider
only interval analysis.

For our experimentation, we used a subset of 200 randomly chosen images from
the Oxford Buildings dataset~\cite{oxfordimages} consisting of 5062 images.
The set of 200 images is partitioned into two equal halves: training and
test sets. The training set is used to obtain estimates of integral
bitwidths at various stages through profiling. The test set is used to
evaluate the effectiveness of the bitwidth estimates obtained for quality
and power. Figure~\ref{fig:hs} shows the average cumulative distribution of
the bitwidth required by the integral part of the pixels in stages $I_x$
and $I_{xy}$  of the HCD benchmark on the training data set.
For example, from Figure~\ref{fig:hs}, we can infer that in stage $I_x$,
95\% of the pixels require less than 5 bits, and all pixels (100\%) can be represented
using 8 bits.
Table~\ref{tab:bitTableHCD} shows the bitwidth estimates obtained from
static and profile-driven analyses for the HCD benchmark.
\begin{table*}[h!]
    \footnotesize
    \caption{Comparison of integral bitwidth estimates using interval, \smtra, 
    and profile-guided analyses for HCD. Fractional bitwidth estimates are also provided in the last row.}
    \vskip 5pt
    \label{tab:bitTableHCD}
\centering
        \begin{tabular}{c | c c c c c c c c c}
        \toprule
        & &&&&HCD&&&&\\
        Stage&Img&I$_x$,I$_y$&I$_{xx}$,I$_{yy}$&I$_{xy}$&S$_{xy}$&S$_{xx}$, S$_{yy}$&det&trace&harris\\
        \midrule
        $\alpha^{\smtra}$ &          8&8&13&14&17&16&33&17&\bf{33}\\
        $\alpha^{IA}$ & 8&8&13&14&17&16&33&17&34\\
        \midrule
        $\alpha^{maxP}$& 8&8&13&14&17&16&30&17&29\\
        $\alpha^{avgP}$& 8&8&13&14&17&16&29&17&29\\
        \midrule
        $\beta$ & 8&5&4&4&3&3&1&1&1       \\
        \bottomrule
        \end{tabular}
\end{table*}
As can be noted from
Table~\ref{tab:bitTableHCD}, the bitwidth estimates from $\alpha^{avgP}$
and $\alpha^{maxP}$  measures are the same for all stages
except for the {\it det} stage.
The estimates from the static analysis techniques match the profile estimates 
except for the {\it det}, {\it trace} and {\it harris} stages. In general, we
expect the profile estimates to be better for stages that occur deeper in
the pipeline. Unlike Optical Flow benchmark, for HCD, \smtra\ analysis 
performs no better than interval analysis except for a single bit 
improvement in stage {\it Harris}. Again, we note that the profile estimates 
also indicate the limit to which the static analysis techniques can be 
improved by using more powerful approaches. Profile information can be 
easily
obtained by executing the HLS C++ program directly without the need for a
heavy weight circuit simulation.

In the next section, we propose a simple and practical greedy search 
algorithm to estimate the number of fractional bits at each stage of the DAG 
while respecting an application specific quality constraint.
%

\subsubsection{Fractional Bits ($\beta$) Analysis}\label{sec:precision}
The number of fractional bits $\beta_i$ required at a stage $i$ depends on the
application and cannot be estimated in an application independent manner 
similar to the integral bitwidth analysis. Estimating the optimal number of
fractional bits at each stage for a given application metric turns out to be a
non-convex optimization problem in most cases and hence we propose a simple
heuristic search technique that requires a very small number of profile 
passes.

In the profiling technique, we fix the number of integral bits required at 
each stage based on static or profile-driven analysis and increase the 
precision
$\beta$ uniformly across all stages. For each value $\beta$, we estimate
the application-specific error metric. For the HCD benchmark, the error
metric is the percentage of misclassified corners when compared to a design
that uses sufficiently long integral and fractional bits. We can reach an
optimal $\beta$ for a given error tolerance via binary search. Then we make
a single pass on the stages of the DAG in  reverse topologically sorted 
order.
At each stage $I$, we do a binary search on the number of fractional bits
required, $\beta_I$, starting from the initial estimate $\beta$ while
retaining the application specific quality requirement. The last row 
of the Table~\ref{tab:bitTableHCD} shows the fractional bits estimated
at each stage of the HCD benchmark. Note that the later stages of the DAG 
require fewer bits than those stages which occur earlier in the DAG. 
This is due to the fact that errors in earlier stages will have a
greater impact as they get propagated to the downstream stages. Further,
our greedy algorithm is optimizing the bitwidths by considering the 
stages in the reverse topologically sorted order. 
\begin{figure}[thbp]
\begin{center}
\usetikzlibrary{calc,trees,positioning,arrows,chains,shapes.geometric,%
                        decorations.pathreplacing,decorations.pathmorphing,shapes,%
                        matrix,shapes.symbols,fit}   
\definecolor{forestgreen}{rgb}{0.0, 0.5, 0.0}

\tikzstyle{decision} = [diamond, draw, fill=blue!20, 
                                    text width=4.5em, text badly centered, node distance=3cm, inner sep=0pt]

\tikzstyle{block} = [rectangle, 
                                                            rounded corners, 
                                                            draw=black, 
                                                            text width=10em, 
                                                            minimum height=1em, 
                                                            text centered]
\tikzstyle{block1} = [rectangle, 
                                                            rounded corners, 
                                                            draw=black, 
                                                            text width=4em, 
                                                            minimum height=3em, 
                                                            text centered]
\tikzstyle{block2} = [rectangle, 
                                                            rounded corners,
                                                            text width=18em, 
                                                            minimum height=3em, 
                                                            text centered] 
\tikzstyle{block4} = [rectangle, 
                                                            rounded corners,
                                                            text width=2em, 
                                                            minimum height=0.5em, 
                                                            minimum width=0.5em, 
                                                            text centered]    

\tikzstyle{fcircle} =[text=black, circle, minimum size=2em]       
\tikzstyle{line} = [draw, -latex',thick, ->]
\tikzstyle{cloud} = [draw, ellipse,fill=red!20, node distance=3cm,
                              minimum height=2em, text width =6em, scale=0.5]
    
\begin{tikzpicture}[auto,scale=0.8, every node/.style={transform shape}]
    \node [block2] (one) {Code specified in PolyMage \\[0.1em] (sobel.py)};
    \node [block, below of=one,yshift=-0.6cm] (two) {PolyMage-HLS compiler};
    \node [block2, below of=two,yshift=-0.6cm] (three) {Untuned HLS code (sobel.cpp) \\[0.1em] (parameterized by {\tt <typ>})};
    \node [block1, below of=three,xshift=-3.0cm,yshift=-0.8cm] (four) {Interval Analysis};
    \node [block1, below of=three,xshift=-1.0cm,yshift=-0.8cm] (five) {Affine Analysis};
    \node [block1, below of=three,xshift=+1.0cm,yshift=-0.8cm] (fivee) {SMT Analysis};
    \node [block1, below of=three,xshift=+3.0cm,yshift=-0.8cm] (six) {Profile Analysis};
    \node [fcircle, right of = four, xshift=-0cm,scale=0.5, font=\bfseries ] (eleven) {or};
    \node [fcircle, right of = four, xshift=+2cm,scale=0.5, font=\bfseries ] (twelve) {or};
    \node [fcircle, right of = four, xshift=+4cm,scale=0.5, font=\bfseries ] (thirteen) {or};
    \node [block,  below of=twelve ,            yshift=-0.8cm,label={[label distance=0cm]50: $\beta$ Analysis}] (seven) {Fractional Bitwidth Search};
    \node [block4,dashed,draw,below of=seven, xshift=-4cm,yshift=1cm,text width=7em,inner sep=1.2pt] (err) {User Error Specification};
    \node [block2, below of=seven,yshift=-0.8cm] (eight) {sobel.cpp \\[0.1em] (with appropriate ($\alpha,\beta$) fixed-point data type chosen for every stage)};
    \node [block, below of=eight,yshift=-0.5cm] (nine) {HLS Compiler};

    \node [draw, dashed, minimum height = 1.2 cm, rounded corners, label={[label distance=0cm]50: $\alpha$ Analysis}, minimum width = 7.9cm, 
                fit={
                            (four)(five)(fivee)(six)
                    }
            ] (ten) {};

    \path [line] (one) -- (two);
    \path [line] (two) -- (three);
    \path [line] (three) -- (ten);
    \path [line] (ten) -- (seven);
    \path [line] (seven) -- (eight);
    \path [line] (eight) -- (nine);
    \path [line] (err) -- (seven);
    \end{tikzpicture}
\end{center}
\caption{Overview of the proposed bitwidth analysis framework.}
\label{fig:flow}
\end{figure}
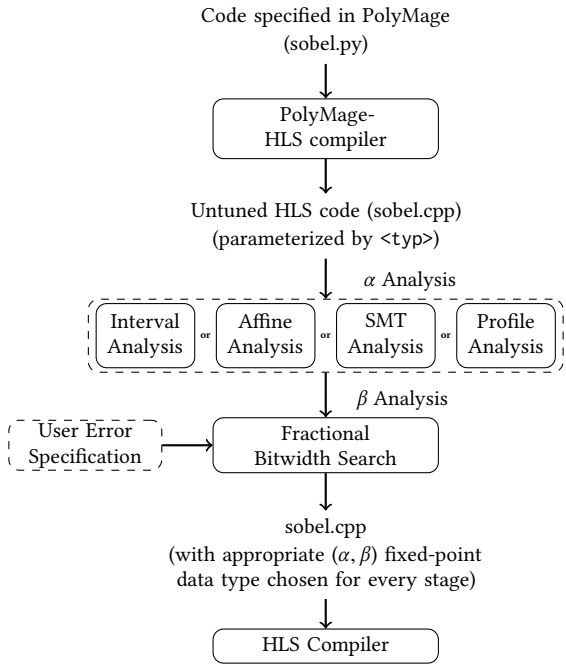
\subsection{Summary of Bitwidth Analysis Framework}
Figure~\ref{fig:flow} summarizes the proposed bitwidth analysis framework. 
We can use the PolyMage-HLS compilation framework first to do a range analysis and 
estimate the integral bitwidths; then use the greedy heuristic to estimate the 
fractional bits required at various stages. For range analysis, we can use one of interval, 
\smtra\ and profile analysis techniques. For interval analysis, the compiler generates
HLS code where the data types of the variables at various stages of the DAG are intervals.
Then the bitwidth estimates are obtained using the intervals obtained by running the HLS 
code. For profile analysis, the compiler generates HLS code wherein the data types of
the variables are of fixed point type with sufficiently large integral and fractional
bitwidths. Then HLS code is run on a sample distribution of input images to arrive at 
integral bitwidth estimates.  For \smtra\ analysis, the compiler generates a 
constraint system which is solved by an SMT solver, such as Z3, to arrive at 
range estimates. 

\section{Experimental Results}\label{sec:experiments}
In this section, we present a detailed area, power and throughput analysis when
variable fixed-point data types are used as against floating-point by 
considering the following four benchmarks: Harris Corner Detection, Unsharp
Mask, Down and Up Sampling, and Optical Flow. Tables~\ref{tab:ofrange},
~\ref{tab:bitTableHCD},~\ref{tab:bitTableUSM} and~\ref{tab:bitTableDUS} show the integral bitwidth 
estimates obtained through interval analysis ($\alpha^{IA}$), \smtra\ analysis
($\alpha^{\smtra}$) and profile analysis ($\alpha^{maxP}$ and $\alpha^{avgP}$); 
and the average fractional bitwidth estimate ($\beta$) obtained through greedy 
heuristic search algorithm. Table~\ref{tab:Allresults} compares the performance
of each benchmark using {\it float} data type and bitwidth estimates obtained 
from different approaches. In these tables, the {\it Quality} column  
corresponds to an application specific quality metric; the {\it Power} 
column gives the power when the design operates at a speed specified in the
adjacent {\it Clk Period} column; {\it latency} columns provide the number 
of
clock cycles required to process an HD image; the next four columns (BRAM,
DSP, FF, LUT, \%slices) summarize area usage; the {\it Min Clk Period}
column gives the maximum frequency of operation for circuit; and the next 
two columns give the throughput and power consumed at the maximum frequency
of operation. Figure~\ref{fig:powersplit} gives the split of power usage by
various components of an FPGA. Unlike the Optical Flow benchmark, the integral 
bitwidth estimates for the benchmarks HCD, USM and DUS using interval and 
\smtra\ analysis techniques is the same. So we do not provide separate area, 
power and throughput analysis for these benchmarks.

We used the Xilinx Zedboard consisting of Zynq-XC7Z020 FPGA device and 
Xilinx Vivado Design Suite 2017.2 version to conduct our experiments. The 
HLS design generated by our PolyMage DSL compiler is synthesized by the 
Vivado HLS compiler. All characteristics are reported post Place and Route.  
We ran C-RTL
co-simulations to generate switching activity (SAIF) file for reporting 
detailed power consumption across the design.

\begin{table*}[t]
    \footnotesize
    \caption{Power, area and throughput analysis for HCD, DUS, USM and OF benchmarks using float and integral bitwidth estimates obtained using
    static and profile-driven analyses. Fractional bitwidths are determined based on the greedy heuristic search approach.}
    \vskip 5pt
    \label{tab:Allresults}
    \centering
    \Ress
\end{table*}

\subsection{Harris Corner Detection}
\begin{figure}[!htb]
    \centering
    \begin{minipage}[t]{0.45\textwidth}
        \centering
\resizebox{\linewidth}{!}{
        \harrisProfPower
        }
        \caption{Error and power variation for the HCD benchmark by fixing the number of integral bits to profile estimated values $\alpha_i^{avgP}$ and varying $\beta$ uniformly across all the stages.}
        \label{fig:uniformbeta}
    \end{minipage}%
    \hfill
    \begin{minipage}[t]{0.45\textwidth}
        \centering
\resizebox{\linewidth}{!}{
        \includegraphics{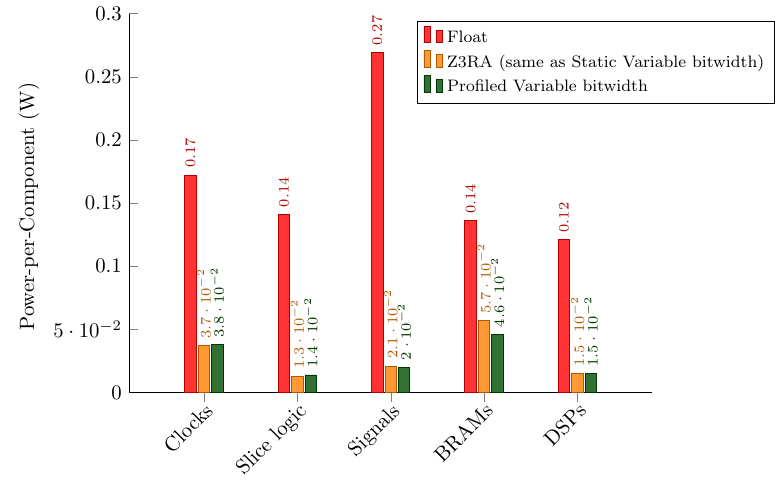}
        }
        \caption{Power consumption by individual components on FPGA for HCD.\label{fig:powersplit}}
    \end{minipage}
\end{figure}

Table~\ref{tab:bitTableHCD} summarizes the integral and fractional bitwidth estimates 
obtained at each stage of the HCD benchmark through various analysis techniques.
The results in this table are commented upon in Sections~\ref{sec:profile} 
and~\ref{sec:precision}. Figure~\ref{fig:uniformbeta} shows the average percentage of 
pixels correctly 
classified by the HCD benchmark on the test image set by varying the 
fractional bits uniformly across all the stages while fixing the integral 
bitwidth estimates obtained via profiling ($\alpha_i^{avgP}$). It also 
contains estimates of power consumption with varying fractional bits for the 
Xilinx ZED FPGA board.
It can be noted from the graph that the fractional bits do not affect the 
accuracy of corner classification, and we thus get more than 99\% accuracy 
even with zero fractional bits. From this graph, we infer that one can 
obtain close to 100\% accuracy by using 8 fractional bits uniformly across 
all the stages. We then make a backward pass on the stages of the
HCD benchmark to drop the fractional bits further without any significant loss
in accuracy and the row corresponding to $\beta$ in 
Table~\ref{tab:bitTableHCD} shows the final fractional bitwidths. Due to space
constraint, we do not provide a graph such as Figure~\ref{fig:uniformbeta} for
the rest of the benchmarks. We can notice from Table~\ref{tab:Allresults} that by 
using bitwidth estimates from interval analysis, we obtain 99.999\% accuracy with
a power consumption of 0.263~W. The power savings are 
3.8$\times$ lower when compared with the floating-point design and 4\% more when 
compared with the profile-estimate based design. The savings
on the percentage of FPGA slices used is around 6.2$\times$. From the last 
3 columns of the table, we can notice that the fixed-point designs can operate
at a higher frequency achieving better throughput while consuming lesser power. 

Figure~\ref{fig:powersplit} shows the detailed power analysis for floating-point and fixed-point design. It shows only the significant components of the dynamic power consumed, and in all the designs, the static power consumption is around 0.122~W.

\begin{figure*}[!htb]
    \begin{minipage}[t]{0.49\linewidth}
        \centering
    \resizebox{0.95\linewidth}{!}{
        \usmDag
    }
    \caption{Pipeline DAG structure for USM benchmark.}
    \label{fig:USMdag}
    \end{minipage}
    \begin{minipage}[t]{0.49\linewidth}
    \centering
    \resizebox{\linewidth}{!}{%
        \begin{tabular}{c | c c c c c }
        \toprule
        Stage&Img&blur$_x$&blur$_y$&sharpen&mask\\
        \midrule
        $\alpha^{Z3RA}$ &8&8&8&10&9 \\
        $\alpha^{IA}$  &8&8&8&10&9 \\
        \midrule
        $\alpha^{maxP}$ &8&8&8&10&9 \\
        $\alpha^{avgP}$ &8&8&8&10&9 \\
        \midrule
        $\beta$  &0&2&3&4&4  \\
        \bottomrule
        \end{tabular}
        }%
        \captionof{table}{Comparison of integral bitwidth estimates using interval, \smtra, and profile-guided analyses for USM. Fractional bitwidth estimates are also provided in the last row.}
        \label{tab:bitTableUSM}
\end{minipage}
\end{figure*}

\subsection{Unsharp Mask (USM)}
The Unsharp Mask (USM) benchmark sharpens an input image and its computational
DAG is provided in Figure~\ref{fig:USMdag}. The input image is blurred across 
x-axis and y-axis by the stencil stages {\it blurx} and 
{\it blury} successively. Then it passes through 
the {\it sharpen} stage, which is a point-wise computation.  Finally, the 
{\it masked} stage compares each pixel from the output of the {\it sharpen} 
stage with a threshold value. Depending on whether the pixel value is less 
than threshold, the corresponding pixel from either the original input image 
or the sharpened image is chosen for output. We highlight an important 
observation here: even if we make an error in computing a pixel value from 
the {\it sharpen} stage, as long as it is less than the threshold, the right 
output pixel is chosen. Based on this observation, we define an error metric 
that is the fraction of pixels that were misclassified in the {\it masked}  
stage due to variable width fixed-point representation as against 
floating-point representation. We define a second quality metric that is the 
root mean squared error between correctly classified pixel values and their 
floating-point counterparts.

Table~\ref{tab:bitTableUSM} shows the integral and fractional bitwidths
required at various stages of the USM benchmark obtained from static (interval
and \smtra) and profile analyses. It can be noted that the estimates obtained by
the static and profile analyses are the same. Table~\ref{tab:Allresults} shows
that there is a factor of 1.6$\times$ improvement in power when compared to the
floating-point design with negligible root mean squared error and classification
error. With respect to the number of FPGA slices used, there is a factor of
2.6$\times$ improvement. Table~\ref{tab:Allresults} also shows the maximum
frequency of operation for each of the designs, the throughput at that level and
power consumption. From the last 3 columns of the table, we can infer that by
operating the fixed-point design at a higher frequency, 6\% increase in
throughput can be achieved while consuming 1.7x lower power.

\begin{figure*}[!htb]
    \begin{minipage}[t]{0.49\linewidth}
    \centering
    \resizebox{0.95\linewidth}{!}{
            \dusDag
    }
    \caption{Pipeline DAG structure for DUS benchmark.}
    \label{fig:DUSdag}
    \end{minipage}
    \begin{minipage}[t]{0.49\linewidth}
    \centering
    \resizebox{0.8\linewidth}{!}{%
        \begin{tabular}{c | c c c c c }
        \toprule
        Stage&Img&D$_x$&D$_y$&U$_x$&U$_y$\\
        \midrule
        $\alpha^{Z3RA}$ &8&8&8&8&8\\
        $\alpha^{IA}$   &8&8&8&8&8\\
        \midrule
        $\alpha^{maxP}$  &8&8&8&8&8\\
        $\alpha^{avgP}$  &8&8&8&8&8\\
        \midrule
        $\beta$    &0&3&6&8&10\\
        \bottomrule
        \end{tabular}
    }
    \captionof{table}{Comparison of integral bitwidth estimates using interval, \smtra, and profile-guided analyses for DUS. Fractional bitwidth estimates are also provided in the last row.}
    \label{tab:bitTableDUS}
    \end{minipage}
\end{figure*}

\subsection{Down and Up Sampling (DUS)}
Down and Up Sampling (DUS) benchmark has a linear DAG structure as shown in Figure~\ref{fig:DUSdag}. The image is first 
downsampled along the $x$-axis in stage D$_x$ and is further downsampled along the $y$-axis in stage D$_y$.
It  is then upsampled again along the $x$ and $y$ axes in the stages $U_x$ and $U_y$ 
respectively. For the sake of conciseness, we avoid including the DUS 
PolyMage code. All four stages comprise stencil computations.

The integral bitwidths estimated by both the interval and \smtra\ analyses is
equal to 8 at all the stages of DUS. We use the same set of training images as
that of HCD benchmark for estimating the integral and fractional bitwidths via
profiling.  The profile estimates yielded the same integral bitwidth requirement
of 8 at all the stages. We use Peak Signal to Noise Ratio (PSNR) as a quality
metric where the reference image is obtained by using a sufficiently wide data
type. We set the required PSNR to infinity and the resulting fractional
bitwidths determined by our greedy precision analyzer is shown in the last row
of the Table~\ref{tab:bitTableDUS}. Table~\ref{tab:Allresults} shows that there
is a factor of 1.7$\times$ reduction in power using tuned fixed-point data types
when compared with using floating-point data type without loss of any accuracy.
With respect to area, there is a 4$\times$ improvement in terms of number of
slices used.   Also, the fixed-point designs use no DSP blocks at all when
compared with floating-point design which uses 54 DSPs. At the peak possible
frequency of operation, fixed-point design achieves 13.6\% increase in
throughput while consuming 1.6x lesser power.

\subsection{Optical Flow (OF)} 
The Optical Flow (OF) benchmark computes the velocity of individual pixels 
from an image frame and its time-shifted version. Our implementation is 
based on  the Horn-Schunck algorithm \cite{horn1981determining} and consists 
of 30 stages.  The first 10 stages are pre-processing stages and the last 20 
stages are obtained by repeating a set of five stages for four times.  The 
accuracy of motion estimation can be improved by repeating  the 5-stage set 
more times.  Optical flow is a heavily used image processing algorithm in 
many computer vision applications.
There have been many efforts in the past to implement optical flow on 
FPGAs~\cite{JavierTCSVT'06,AlanISSC'10,EnZhuTCSVT'16} for power and 
performance benefits.

Table~\ref{tab:ofrange} shows the estimated integral bitwidths required 
at various stages of the Optical Flow benchmark. We notice that for stages 
deeper in the pipeline, the difference between estimates obtained via interval  
analysis and profiling are substantial. The profile estimates are obtained 
from a training data set and for testing purpose, we use RubberWhale and 
Dimetrodon image sequences from the Middlebury dataset~\cite{MBuryDataSet}.  
Section~\ref{sec:smt} provides a detailed discussion on this and shows how 
the \smtra\ analysis can overcome the inadequacies of the interval arithmetic
based analyses techniques and gives estimates which almost match profile estimates.
For computing the accuracy, we use the Average Angular Error (AAE) metric as 
discussed in \cite{Fleet:1990},\cite{Otte:1994}. The reference motion 
vectors are obtained by using sufficiently wide fixed-point data types at 
all stages. 

It can be noticed from Table~\ref{tab:Allresults} that by  using bitwidth
estimates from Z3RA analysis, we obtain similar accuracy as  profile-driven
analysis with a power consumption of 0.328~W. The power savings are  1.9$\times$
lower when compared with the floating-point design and 5.4\% more when  compared
with the profile-estimate based design. The savings on the percentage of FPGA
slices used is around 2.5$\times$.  From the last 3 columns of the table, we can
notice that the Z3RA fixed-point design can operate at a higher frequency
achieving 25\% more throughput than the floating point design while consuming
lesser power.

\section{Conclusions}
\label{sec:conclusions}
The input, output and intermediate values generated in many image processing
applications have a limited range. Furthermore, these  applications are
resilient to errors arising from factors such as limited  precision
representation, inaccurate computations, and other potential noise  sources. In
this work, we exploited these properties to generate power and  area-efficient
hardware designs for a given image processing pipeline by  using custom
fixed-point data types at various stages. We showed that domain-specific 
languages facilitate the application of interval and affine arithmetic 
analyses on larger benchmarks with ease. Further, we proposed a new 
range analysis technique using SMT solvers, which overcomes the inherent 
limitations in conventional interval/affine arithmetic techniques, when 
applied to iterative algorithms. The proposed SMT solver based range 
analysis technique also uses the DSL specification of the program to reduce 
the number of constraints and variables in the constraint system, thereby 
making it a feasible technique to adopt in practice. Then, we compared the 
effectiveness of the static analysis techniques against a profile-driven 
approach that automatically takes into account properties of input image 
distribution and any correlation between computations on spatially proximal 
pixels. In addition, the analysis revealed the limit of possible improvement 
for any static analysis technique for integral bitwidth estimation.  
Finally, to estimate the number of fractional bits, we used uniform 
bitwidths across all the stages of the pipeline, and then used a simple 
greedy search to arrive at a suitable bitwidth at each stage while 
satisfying an application-specific quality criterion. Overall, the results 
effectively demonstrate how information exposed through a high-level DSL 
approach could be exploited in practical fixed-point data type analysis 
techniques and to perform detailed impact studies on much larger image 
processing pipelines than previously studied.

\section{Acknowledgments}
\label{sec:acknowledgments}
We would like to gratefully acknowledge the Science and Engineering Research Board (SERB), Government of India for funded this research work in part through a grant under its EMR program (EMR/2016/008015).

\bibliographystyle{ACM-Reference-Format}
\bibliography{bibfile}

\end{document}